\documentclass{aa}  
\usepackage{graphicx}
\usepackage{txfonts}
\usepackage{color}
\usepackage{sidecap}
\newcommand{\kms}{km\,s$^{-1}$}

\newcommand{\um}{$\mu$m}

\begin{document} 

   \title{A radical transition in the post-main-sequence system U Equulei}

   \author{Tomek Kami\'nski\inst{1}
          \and Mirek R. Schmidt\inst{1}
          \and Anlaug Amanda Djupvik\inst{2,3} 
          \and Karl M.\ Menten\inst{4} 
          \and Alex Kraus\inst{4}
          \and Krystian Iłkiewicz\inst{5}
          \and Thomas Steinmetz\inst{1}
          \and Muhammad Zain Mobeen\inst{1}
          \and Ryszard Szczerba\inst{1}
          %\fnmsep\thanks{Just to show the usage
          %of the elements in the author field}
          }
\institute{\centering
Nicolaus Copernicus Astronomical Center, Polish Academy of Sciences, Rabia{\'n}ska 8, 87-100 Toru{\'n}, Poland, \\\email{tomkam@ncac.torun.pl}\label{inst1}
\and Nordic Optical Telescope, Rambla Jos\'e Ana Fern\'andez P\'erez 7, 38711 Bre\~na Baja, Spain \label{inst3}
\and Department of Physics and Astronomy, Aarhus University, Ny Munkegade 120, 8000 Aarhus C, Denmark \label{inst4}
\and Max-Planck-Institut f\"ur Radioastronomie, Auf dem H\"ugel 69, 53-121 Bonn, Germany \label{inst4}
\and Astronomical Observatory, University of Warsaw, Al. Ujazdowskie 4, 00-478 Warszawa, Poland \label{inst5}
}
%\date{Received March, 2020}
\authorrunning{T. Kami\'nski et al.}
\titlerunning{A transition in U\,Equ}

\abstract
  % context heading (optional)
  % {} leave it empty if necessary  
   {U Equ is an unusual maser-hosting infrared source discovered in the 1990s. It was tentatively classified as a post-AGB star with a unique optical spectrum displaying rare emission and absorption features from molecular gas at a temperature of about 500\,K. In 2022, we serendipitously discovered that its optical spectrum has drastically changed since the last observations in the 1990s.}
  % aims heading (mandatory)
   {We aim to characterize the drastic change in the spectrum and analyze the photometric behavior of the object since 1989. }
  % methods heading (mandatory)
   {Optical high-resolution spectra of U\,Equ from the Southern African Large Telescope are supplemented by archival data and near-infrared photometry from the Nordic Optical Telescope. New spectral line observations with the Effelsberg 100\,m radio telescope and ALMA are presented. Radiative transfer modeling of multiple epoch spectral energy distributions is performed.}
  % results heading (mandatory)
   {No circumstellar molecular features are present in the contemporary optical spectra of U\,Equ. Non-photospheric absorption and emission from neutral and ionized species dominate the current spectrum. Some of the observed features indicate an outflow with a projected terminal velocity of 215\,\kms. Broad H\&K lines of [\ion{Ca}{II}] indicate a photosphere of spectral type F or similar. For the first time, we find SiO $J=$1--0 $\varv$=1 maser emission in U\,Equ. Our collected photometric measurements show that the source has been monotonically increasing its optical and near-IR fluxes since about the beginning of this century and continues to do so. The current rise in the optical regime is about 1 mag. Spectral energy distributions at different epochs show the presence of dusty circumstellar material that is very likely arranged in a highly-inclined disk. Adopting a distance of 4\,kpc, informed by the Gaia parallax of U\,Equ, we find that the source's luminosity is on the order of 10$^4$ L$_{\sun}$. This luminosity has likely increased by a factor of a few in the last decades, which is most probably related to the drastic change in the optical circumstellar spectrum of the object.}
   %Recombination lines of hydrogen and the infrared [\ion{Ca}{II}] triplet are strongest observed features but numerous weaker lines of neutral and ionized metals, including Ti, Fe, and Cr, are defining the optical spectrum. We find evidence of a high-velocity (300\,\kms) outflow whose spectral variations are observed on a time scale of months. Archival photometry of the objects shows slow but a steady rise in visual and IR fluxes over the last 2--3 decades. The source could possibly have changed it luminosity by a factor of a few.}
  % conclusions heading (optional), leave it empty if necessary   
   {The object has changed considerably in the last three decades either due to geometrical reconfiguration of the circumstellar medium, evolutionary changes in the central star, or owing to an accretion event that has  
   started in the system very recently. Observationally, U\,Equ appears to resemble the Category\,0 of disk-hosting post-AGB stars of \cite{KluskaReview}, especially the post-common envelope binary HD\,101584. It is uncertain if the drastic spectral change and the associated optical/mid-IR rise in brightness witnessed in U\,Equ are common in post-AGB stars but such a radical change may be related to the real-time onset of the evolution of the system into a planetary nebula. We find that the post-AGB star V576\,Car has undergone a similar transformation as U\,Equ in the last few decades, so the phenomenon is not extremely rare.}  
   %Although the amplitude of the photometric changes is not significant, the observed properties of the objects are drastically different. Based on the remarkable similarity to a few objcets thought to have recently ondergone common-envelope ejection, we speculate that we could have witnessed envelope ejection in U\,Equ in real time.}


\maketitle
%
%-------------------------------------------------------------------
\section{Introduction}
The transition of an asymptotic giant branch (AGB) star into a white dwarf (WD) is a relatively poorly known process. The post-AGB (or pre-planetary nebula; PPN) phase is expected to last on the order of 10 to 10$^5$ years \citep{schoen83,Miller}, during which the star is thought to shed most of its outer layers and expose them to the harsh radiation of its inner core, which, with time, will cool down to become a proto-WD \citep{paczynski}. Although this phase is short compared to the main-sequence and the AGB timespan, the number of evolved AGB stars in the Galaxy is big enough \citep[e.g.,][]{Iwanek} for us to be able to catch some objects in this transition stage. The hallmark of this evolutionary shift, as predicted by the theory, should be a steep increase in the  effective temperature of the photosphere, occurring at a nearly constant luminosity. However, so far, we have not identified firmly such objects in our observations, especially at the onset of the post-AGB evolution when the star is still embedded in a cloud consisting of dust and  atomic/molecular gas accumulated during the turbulent and mass-reducing AGB phase. Although the post-AGB star with increasing effective temperature ($T_{\rm eff}$) cannot yet fully photoionize this circumstellar envelope (which only happens in the planetary nebula, PN, phase) the evolution off the AGB should have a significant impact on that environment. The effect on the circumstellar medium may be actually easier to spot than the direct change in the photospheric temperature. In practice, the circumstellar environments of stars classified as post-AGB display an enormous level of complexity \citep[cf.][]{oh231}, often caused by interactions with a companion, so that genuine changes caused by the actual evolution of the star may not be easily recognizable.   

%Many, possibly even a third, of post-AGB stars have a companion, and some may host planets \citep{KluskaReview}. The binarity is often associated with the presence of a dusty disk, which is relatively easily recognizable in systems' spectral energy distribution (SED). A large sample of those disks were imaged with interferometric techniques \citep[][and references therein]{KluskaReview}. Acrreation of the circumbinary disk material into the binary cause striking chemical peculiarities in the atmospheres of the post-AGB components.  

U\,Equ \citep{barnbaum} and V576\,Car \citep{couch} must be amongst the most bizarre objects classified as post-AGB stars. When observed at optical wavelengths in the 1990s, their spectra displayed strong molecular bands in absorption (in both objects) or in both emission and absorption (U\,Equ). The bands arise in warm ($\lesssim$1000\,K) molecular gas located relatively close to the stars, which, along our line of sight, are strongly obscured by dust. Despite the severe obscuration, the main stellar component in both objects was classified as types F--K which are too hot to be typically associated with circumstellar environments rich in molecules, especially close to the stellar photospheres. Indeed, U\,Equ and V576\,Car have long been considered very unusual. We focus here especially on U\,Equ, which showed molecular electronic bands of various metal oxides in emission, most prominently from TiO and VO. This is a very rare spectroscopic feature, and only a handful of other objects have been reported to possess such emission \citep{lloyd}. Red nova remnants, such as V4332\,Sgr, V1309\,Sco, and V838\,Mon, have shown this phenomenon \citep{KamiV4332,KamiV1309,KamiV838} for years to decades after their outbursts. Several young stellar objects displayed such molecular emission features, too, in relation to an accretion-driven outburst \citep[e.g.,][]{ysos} but it is typically a short-lived feature in the optical spectra. The red supergiant VY\,CMa is the only known source whose optical spectrum has molecular emission lines permanently in emission, at least since its first spectroscopic observations \citep{wallerstein1971,herbig74,emTiO,KamiAlO,humphreys}. U\,Equ is sometimes singled out as an example of a non-eruptive star with molecular emission bands.  

\begin{figure*}
    \centering \sidecaption
   \includegraphics[width=12cm]{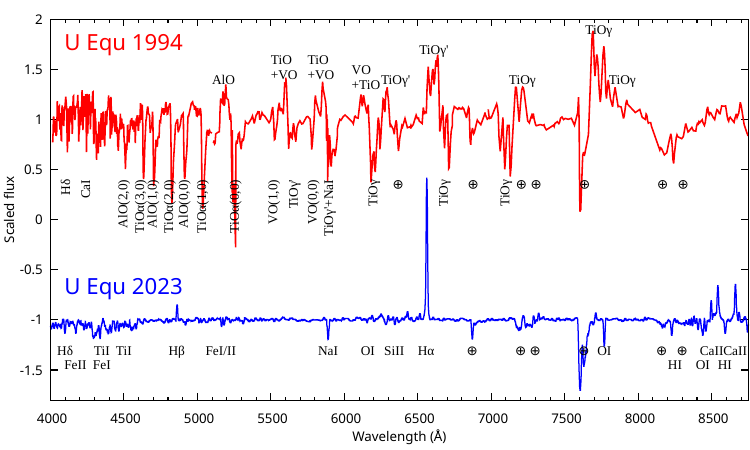}
    \caption{Spectra of U\,Equ obtained in 1994 \citep[top; extracted from][]{barnbaum} and in 2022--2023 (bottom). The spectrum had completely changed the appearance from being dominated by molecular absorption-emission bands and requiring low excitation temperatures into a circumstellar spectrum of partially ionized atomic gas at a much higher excitation temperature. In this paper, we describe and try to interpret this dramatic change in U\,Equ. The bottom spectrum was smoothed to match the resolution of the earlier observations. The earth symbols mark positions of strongest telluric features.}
    \label{fig-first}
\end{figure*}

To investigate the phenomenon of molecular line emission in different types of stars, we recently included U\,Equ in a survey in which we acquire high-resolution spectra  with the HRS 
instrument\footnote{\url{https://astronomers.salt.ac.za/instruments/hrs/}} at the Southern African Large Telescope (SALT). To our great surprise, the optical spectra acquired in 2022 and 2023 for U\,Equ showed a completely different appearance from those described in the literature. This dramatic change is illustrated in Fig.\,\ref{fig-first} where we compare the recent SALT spectrum to one obtained in 1994. All molecular features have disappeared and were replaced by numerous atomic lines in absorption and emission, and arising from gas of low ionization but at temperatures much higher than required for the historical circumstellar spectra of U\,Equ. Here, we investigate the nature and cause of this remarkable transformation of U\,Equ. In Sect.\,\ref{sec-history}, we review what has been known about U\,Equ before our observations. In Sect.\,\ref{sec-photometry}, we present photometric data that allow us to understand the object's behavior over time scales of decades. We analyze and discuss these changes in Sect.\,\ref{sect-phot-analysis}. Constraints on the physical characteristics of U\,Equ put by Gaia observations are briefly presented in Sect.\,\ref{sec-gaia}. Complementary observations of molecular species at millimeter and radio wavelengths are presented in Sects.\,\ref{sect-alma} and \ref{sect-sio}, respectively. The SALT observations are described in Sect.\,\ref{sect-salt-obs} and the new spectra are presented and analyzed in Sect.\,\ref{sec-salt-anal}. We discuss the essence of the observed changes and nature of U\,Equ in Sect.\,\ref{sect-discuss}. Therein, in Sect.\,\ref{V576Car}, we also mention observations of the other post-AGB star rich in optical molecular line absorption, V576\,Car, whose spectrum and photometric history have undergone similar changes as U\,Equ.

%-------------------------------------------------------------------
\section{Prior observations and impressions on U\,Equ}\label{sec-history}
Although it was first reported by \citet{reinmuth} in a list of variable stars, the modern history of U\,Equ can be traced back to the late 1980s, when far-infrared (FIR) source that had been detected by the Infrared Astronomical Satellite (IRAS\,20547+0247) was recognized as a host of OH and H$_2$O masers. The maser observations of U\,Equ are described separately in Sect.\,\ref{sec-masers}. In addition to photometric measurements, IRAS took a spectrum of U\,Equ with the Low Resolution Spectrograph in the 8--22\,\um\ range \citep{iras-lrs}. The spectrum was described as having a strong 10\,\um\ silicate feature. As a bright infrared (IR) source with IRAS fluxes (epoch 1983.5) of 45.6, 33.7, 9.98, and 2.7\,Jy at 12, 25, 60, and 100\,\um, respectively, it was often classified as an OH/IR star. However, several authors noticed that its location well below the Galactic plane, with $l=-$26\fdg11, is rather unusual for this class of stars and might indicate a halo object. 

\citet[][hereafter BOM96]{barnbaum} brought U\,Equ to spotlight when they acquired first optical spectra of the object in 1994. Their medium-resolution spectra showed a multitude of absorption and emission features, the majority of which were assigned to electronic bands of TiO, VO, and AlO. While the emission bands most likely were produced by fluorescence, the absorption spectrum was non-photospheric, given the lines' narrow width and low inferred excitation temperatures, that is with $T_{\rm ex}$=700--1500\,K (most of the bands were from transitions related to the ground electronic level and low vibrational states). Such a spectrum appeared very unusual to the observers at that time \citep[cf.][]{lloyd}.

Based on the observation of the 10\,\um\ silicate feature in absorption and on IRAS colors, BOM96 postulated that U\,Equ has a circumstellar disk seen edge-on and that this dusty disk was obscuring the star along our line of sight. The optical molecular features were interpreted as arising in the inner parts of the circumstellar medium, but BOM96 did not locate them in the disk itself. 

Based on that data, little could be inferred on the photospheric spectrum of U\,Equ, as it was strongly contaminated by the omnipresent circumstellar features. 
%In passing, \cite{lloyd} mentioned the stellar spectrum may be of type G or K, but since the source of this observation is unknown, it should be treated with caution. Although it is nowhere stated directly, in the interpretation of BOM96 -- with the star being completely obscured by a disk -- the photospheric spectrum can still be seen through light scattered above and below the disk plane. Such a spectrum was likely highly polarized and attenuated by circumstellar extinction. 
Based on the highly contaminated spectral features present in the 1994 spectra, BOM96 proposed the central star to be a giant (luminosity class III) with a spectral type of mid G to early K. This was mainly based on a few absorption lines of hydrogen and of a few metals. We consider these estimates highly uncertain, based on the nature and quality of their spectra. The radial velocity of the photospheric features could not be measured reliably in the spectra, too. 

BOM96 also combined photometric observations from different epochs to construct a spectral energy distribution (SED) of the system. While an FIR continuum excess, traced by the emission in the IRAS bands, indicated warm dust at a temperature of about 350\,K, the intrinsic stellar spectrum and extinction components are not easily readable from their data. With their preferred spectral type, however, BOM96 estimate the circumstellar reddening, $E(B-V)$, as 0.68\,mag (which corresponds to a visual extinction, $A_V$ of 2.1\,mag); it is larger than their adopted interstellar reddening of $E(B-V)$=0.09\,mag. Assuming a luminosity of 10$^3$\,L$_{\sun}$, from the SED they obtained a distance of 1.5\,kpc. BOM96 concluded that a dust-obscured evolved giant was the most likely identification of the star, especially given the presence of OH and H$_2$O masers. The high Galactic latitude of the object, in their opinion, excluded a young stellar object.

\cite{Geballe} observed U\,Equ spectroscopically in the near-IR in 1997--2003. They found an absorption features of CO and H$_2$O  which arise at temperatures of 500--1000\,K, similar to the temperature of the gas probed by the optical bands of metal oxides. Resolved lines had a full width half maximum of 35\,\kms\ and were centered at --91\,\kms\ in the heliocentric rest frame (--78\,\kms\ in the LSR). A weak ro-vibrational absorption line $\varv$=1--0 $S(1)$ of H$_2$ was also observed at a similar velocity. Line ratios between the $^{12}$CO to $^{13}$CO isotopologues suggest a $^{12}$C/$^{13}$C ratio of 4, but due to opacity effects it may be significantly higher. Indeed, the value of 4 would be very perplexing in an O-rich object. \cite{Geballe} estimated column densities for the three molecular species. The values they obtain (e.g., $2\times 10^{20}$\,cm$^{-2}$ for CO) appear to be too low compared to what one would infer from the circumstellar extinction of $A_V =2.1$\,mag assumed by BOM96. One possible interpretation is an extremely low dust content in the circumstellar material -- which would be contradictory to the FIR excess -- but the uncertainties in both quantities are very large. The central velocity of absorption lines was interpreted by \citet{Geballe} as representing that of the stellar photosphere. However, we note that the width and the excitation temperature characterizing these lines readily imply a  circumstellar origin. 
%As is discussed later on, this velocity is highly blueshifted with respect to the masers and to the pure rotational CO emission (Sect.\,\ref{sect-alma}), which we interpret as a signature of an outflow seen against an extended (pseudo-)photosphere. 

Several imaging experiments in the visual and infrared, including one with the Hubble Space Telescope (HST) and Very Large Telescope (VLT), did not spatially resolve the U\,Equ system \citep{barnbaum,Geballe,siodmiak,lagadec}.

%-----------------------------------------------------------------------
\begin{figure*}
    %\centering 
    %\sidecaption
%   \includegraphics[width=12cm]{plotMasers-crop.pdf}
   \includegraphics[trim=9 3 0 0, width=12cm]{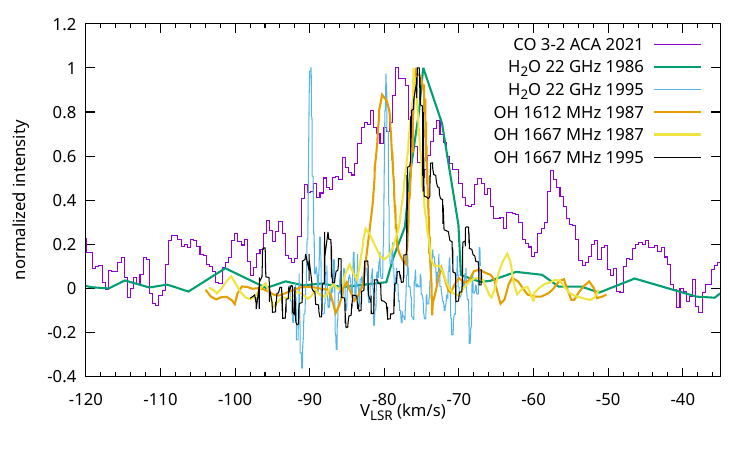}
   \includegraphics[scale=1.0]{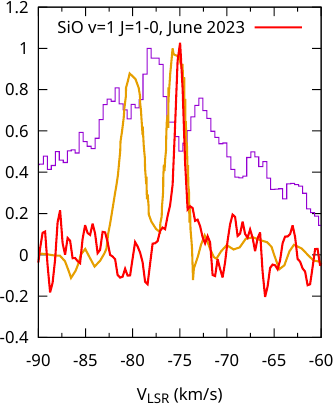}
   \includegraphics[width=11.5cm]{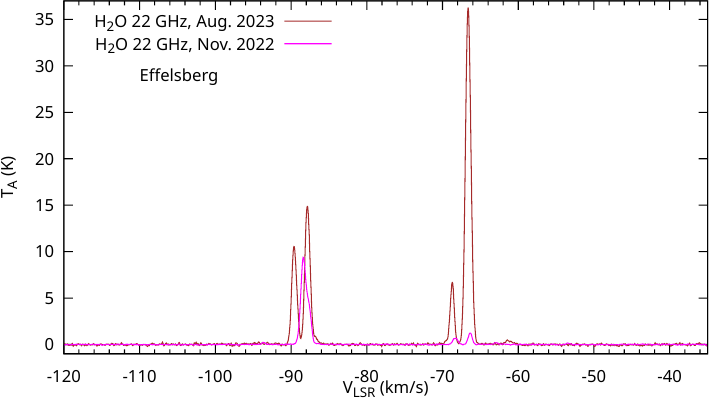}
    \caption{{\bf Top left:} Maser emission of OH and H$_2$O extracted from the literature data is compared to the CO $J$=3--2 profile obtained more recently with ALMA. The spectra were acquired with different spectral resolutions and are shown here normalized to the peak. 
    {\bf Right:} On a reduced LSR velocity axis, the spectrum of the SiO $J$=1--0 $\varv$=1 maser (red) is compared to that of the 1667 MHz OH maser (observed in 1987, orange) and to the CO $J$=3--2 line (purple).
    {\bf Bottom:} Spectra of the water masers acquired recently with the Effelsberg telescope. They are shown in the absolute intensity scale to demonstrate the variability in line fluxes. 
    The velocity in the heliocentric rest frame is 13.1\,\kms\ lower than in the LSR frame adopted here.}
    \label{fig-masers1}
\end{figure*}

\subsection{Masers in the OH/IR phase}\label{sec-masers}
Masers in U\,Equ has been long thought to be very unusual. The H$_2$O and OH masers have been observed in U\,Equ at least since 1985 \citep{firstMasers} and have been known to be variable in intensity and velocity. While variability is a common phenomenon for maser lines, that observed toward U\,Equ was quite remarkable. The first observations of the H$_2$O 22 GHz masers revealed a single-peak spectral feature at $V_{\rm LSR}$=--75.0\,\kms. However, later observations in 1995 showed two weaker features, at --79.9 and --90.0\,\kms, demonstrating behavior untypical for a circumstellar maser source. The OH masers  at lower frequencies have been observed much more often. In May 1987, the 1612 and 1667\,MHz OH masers were observed at LSR velocities of --80.4 and --75.7\,\kms\ \citep{OHfirst}, but only the 1612\,MHz feature was clearly double. The OH and H$_2$O lines showed little overlap in velocities, and only one component had a similar velocity in both species. Two components of 1612\,MHz OH masers were also detected in August 1985 by \citet{Chengalur} at --79.4 and --75.3\,\kms\ (LSR). \cite{barnbaum} observed the OH\,1667 MHz masers in 1995 and noticed profile changes on time scales of months. No double structure, known from the earlier epochs, was found. \cite{Etoka} observed both OH transitions around the same time (in 1994) and found only single-peaked features, as well. In the 1994--1995 observations, the OH profile was centered near --75.7\,\kms, with an error of about a few \kms\ (the errors are not indicated in the original papers). In those observations, the OH masers were getting weaker. The 1612\,MHz emission was not detected in 1995. In a survey by \cite{deadStars}, the same OH transition was targeted but not detected in 1999--2001. This led B. M. Lewis to conclude that U\,Equ (= OH\,051.3+26.1) is an example of a `dead OH/IR star', a fate he predicted for all OH/IR stars on timescales of $\sim$1700\,yr.  Based only on the OH maser observations, it is clear that the object was experiencing a radical change at least since the beginning of the 1990s. We compare the literature  reports on the spectral positions OH and H$_2$O maser lines in Fig.\,\ref{fig-masers1}.

No SiO masers had ever been reported in U\,Equ. \cite{Deguchi} put only upper limits with an rms of 0.06\,Jy for the $J$=1--0 rotational lines in the  $\varv$=1 and 2 vibrationally-excited states in 2001. Any earlier reports would be valuable, but we were not able to find any. Post-AGB objects are almost never associated with SiO masers \citep{engels2001}, but note the prominent case of the pre-planetary nebula OH231.8+4.2 containing a Mira component \citep{Morris1987}.

%to that of CO as seen by the ACA (Fig.\,\ref{fig-masers1}). Although all reported maser features overlap with the broad CO profile (representing the entire source), they are not aligned with the central CO velocity, nor they are distributed symmetrically around that velocity. This stands in stark contrast to many circumstellar masers, including those of evolved stars. 

%Compare to water fountains? The spread in water maser velocities in U\,Equ is smaller than in the so-called  water fountains. 

%---------------------------------------------------------------
\section{Archival photometry}\label{sec-photometry}
To understand the transition that occurred in the optical spectrum of U\,Equ, we tried to reproduce its multiband light curves from all available sources. Below, we summarize and critically evaluate the archival materials that we found. The light curves are shown in Fig.\,\ref{fig-lc}

\subsection{Visual}
Heidelberg-K\"onigstuhl plates reported by \cite{reinmuth} provide two measurements of a source at the location of U\,Equ (the coordinates of his source number 95 converted to J2000 are RA=20:57:16.31 and Dec=02:58:44.78, that is within 2\arcsec\ from the current position). Magnitudes of <15.5 in 1902, and of 14.5 in 1904 and 1915 were measured. These measurements are from unfiltered observations with blue photographic plates.

More Heidelberg-K\"onigstuhl blue plates were scanned and assigned an astrometric solution. From those available at the Heidelberg Digitized Astronomical Plates service{\footnote {\url{https://dc.zah.uni-heidelberg.de/lswscans/res/positions/q/form}}}, we selected ones in which U\,Equ and nearby fields stars are readily seen. Those digitalized plates are from 1902, 1904, 1915, and 1927, and do not show any considerable variation (within photometric accuracy) in the brightness of U\,Equ compared to the field stars (see Fig.\,\ref{fig-plates}). On those dates, and likely in the entire 1902--1927 period, the star did not experience any sudden flux increase by several magnitudes. It is reasonable to assume that the magnitudes stayed close to those reported by Reinmuth, that is near 14.5 mag, in the first quarter of the 20th century. Some variability was however noticed since the star received a variable star designation.
%KMM: One wonders on what basis Reinmuth picked our U Equ as a variable...

Plates of the first Palomar Observatory-National Geographic Sky Survey (POSS) provide two measurements in 1951, 16.05\,mag on a red E plate and 21.77\,mag on a  blue O plate. These magnitudes were measured by \cite{POSS}. The second POSS survey yielded a red magnitude of 13.86 in Sept. 1990, a blue magnitude of 15.88 in June 1993, and $I$=12.41 in August 1994 \citep{USNO}. The same catalog specifies the first-epoch red magnitude as 15.28 and not 16.05 mag. We interpret this discrepancy as a result of intrinsic uncertainties in photometry from these plates. Based on these measurements, the photographic magnitudes did not change by more than 1 mag between 1951 and 1990. 
%KMM I don't understand: E: 15.28 vs. 13.86 mag and O: 21.77 vs. 15.88 Mag. >> 1 mag change! ???

\begin{figure*}
    \centering
   \includegraphics[width=\textwidth]{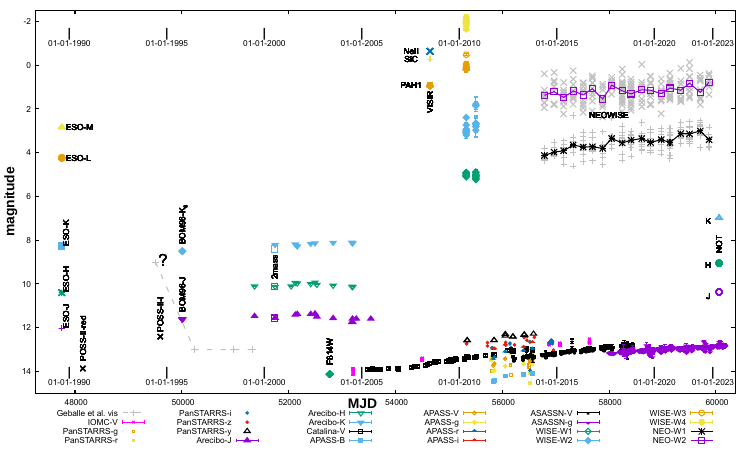}
    \caption{Photometric evolution of U\,Equ.}
    \label{fig-lc}
\end{figure*}

According to \cite{Geballe}, the visual magnitude of U\,Equ in 1994 was 9. No source of this measurement is given, and we find it very unlikely the star had risen to such a low magnitude, especially given the photometric constraints from the second POSS survey. Geballe at al. give also a magnitude 13 for 1996, 1998, and 1999, but again with no details of how those measurements were acquired. We include these photometric points in Fig.\,\ref{fig-lc} but consider them very doubtful.

A $V$ band light curve spanning from 2004 to 2016 was extracted from the INTEGRAL Optical Monitoring Camera (IOMC) data\footnote{\url{https://sdc.cab.inta-csic.es/omc/secure/form_busqueda.jsp?resetForm=true}}. We utilized their $V3$ measurements.

We extracted another set of visual magnitudes from the Catalina Survey\footnote{\url{http://nesssi.cacr.caltech.edu/DataRelease/}} \citep{catalina} for the period 2005--2013. These are unfiltered magnitudes very roughly transformed to $V$.

Several measurements in the $grizy$ bands from 2009--2014 were found in the detection catalog of Pan-STARRS1\footnote{\url{https://catalogs.mast.stsci.edu/panstarrs/}} \citep{pannstarrs}. In a similar time period, measurements were taken by the AAVSO Photometric All-Sky Survey (APASS)\footnote{\url{https://www.aavso.org/apass}}, mainly in the $BVgri$  bands. %We show only selected bands in Fig.\,\ref{fig-lc}.

The ongoing Sky Patrol of the All-Sky Automated Survey for Supernovae (ASASSN)\footnote{\url{https://asas-sn.osu.edu/}} has been observing U\,Equ in $Vg$ bands since 2012 \citep{asassn1,asassn2}. We include here data obtained till June 2023.

%We also extracted photometric sequences from the TESS mission. We used the ASASSN data to set relative scaling between different TESS photometric sequences.

Several  single-epoch measurements in various bands were found. We list them here for completeness, but not all are included in our analysis. 
\begin{itemize}
   \item A Johnson $B$ magnitude of 15.75\,mag was derived from the Ukrainian glass archive \citep{UkrainePlates}\footnote{\url{https://cdsarc.cds.unistra.fr/viz-bin/cat/I/342#/description}} for July 1991. This value is close to that from the blue plate POSS-II measurement of 1993.

  \item Gaia DR3 measurements of $G$=12.730$\pm$0.004, $BP$=13.184$\pm$0.011, and $RP$=12.091$\pm$0.010\,mag. The exact epochs of these measurements are unknown. 

  \item The AAVSO archive has a single $V$ measurement of 12.5\,mag for Nov. 2021.

  \item \cite{siodmiak} provided HST magnitudes of <14.94 and 14.12 in the $F606W$ and $F814W$ filters based on observations from May 2003. The $F606W$ image was saturated.

  \item Public data of the TESS mission are available from Aug. and Sept. 2022. We extracted the data using the {\tt Lightkurve} package \citep{2018ascl.soft12013L} and scaled the raw instrumental intensity to the corresponding observational data from ASASSN. The TESS data have high cadence and show better some short-term variability of the source.

\end{itemize}

Multiband visual colors are mainly available for the period 2010--2013. No colors are essentially known prior to 2005.
The archival plates, although they only sparsely probe the photometric behavior, seem so indicate unchanged magnitudes in the blue and red. If there were any variations, they must have been lower than about 1 mag between 1904 and 1994. Of course, it is always possible that significant changes occurred in the large windows when no one observed U\,Equ.

Measurement uncertainties were rarely given in the above quoted sources. The combined photometric data are shown in Fig.\,\ref{fig-lc}.

%---------------------------------------------------------------------------------------
\subsection{Near-infrared}
A collection of $JHK$ measurements between 1999 and 2005 were taken from the Optical and Near Infrared Atlas of the Arecibo Sample of OH/IR Stars\footnote{\url{https://arecibo.cab.inta-csic.es/}} \citep{AreciboIR}.  Hereafter and in Fig.\,\ref{fig-lc}, we call these points the Arecibo data.

\cite{Fouque} performed $JHKLM$ photometry of the object in Aug. 1989 using the ESO 1\,m telescope. % The useful reddest bands showed a source with $L$=4.24 and $M$=2.85\,mag.

BOM96 presented measurements in the $J$ and $K_s$ bands for Oct. 1995 and obtained with the Shane telescope. %IRAS fluxes were also reported in BOM96 for epoch 1983.5.

The 2MASS measurements in $JHK_S$ were obtained in July 2000 \citep{2mass}. 

\cite{lagadec} used the VISIR instrument on the ESO Very Large Telescope (VLT)  on 29 June 2008 to image U\,Equ at 8.59, 11.85, 12.81\,\um\, that is in the {\tt PAH1}, {\tt SiC}, and {\tt NeII} filters, respectively. We transformed the measured flux densities of 21.3, 35.7, and 41.7\,Jy, respectively, into Vega system magnitudes of 0.94, --0.27, --0.63 in those respective filters.

Observations in all four WISE bands (i.e., at $W1$=3.4, $W2$=4.6, $W3$=12, and, $W4$=22\,\um) were extracted from the AllWISE Multiepoch Photometry Table\footnote{\url{https://irsa.ipac.caltech.edu/cgi-bin/Gator/nph-scan?submit=Select&projshort=WISE}} \citep{wise}. These observations were made in May and Nov. 2010. 

A much longer time sequence of photometric measurements at 3.4 and 4.6\,\um\ was extracted from the NEOWISE Reactivation Single Exposure (L1b) Source Table \citep{neowise}. The $W1$ and $W2$ data were
taken over the period from 2014 May 2014 to Oct. 2022. For better clarity, in Fig.\,\ref{fig-lc} we rebinned the data by averaging measurements obtained within 7 days (original data are shown in gray).

The most recent $JHK_S$ photometry is not archival, but was obtained by ourselves on 25 April 2023 with the NOTCam instrument \citep{notcam} at the Nordic Optical Telescope \citep[NOT;][]{notTel}. The seeing during the observation was of about 0\farcs8. 
Owing to the relatively high brightness of the target, the high-resolution (HR) camera of NOTCam (0\farcs079/pixel) was used and exposure times kept short: 30\,s for the $J$-band, 10.8\,s for the $H$-band, and 3.6\,s for the $K_S$-band. Moreover, for the $K_S$-band, the telescope was slightly defocused to avoid saturation. The images were obtained by dithering in a 3$\times$3 pattern, and reduced with the IRAF package {\tt notcam.cl} to correct for non-linearity, flat-field, background sky, and to finally align, shift, and calculate median combined images. A nearby field with six 2MASS stars, observed right after U\,Equ with the same set-up, was used for the photometric calibration. We converted the 2MASS photometry of these stars to the Maunakea Observatory (MKO) system using the transformations in \citet{Leggett}.  Aperture photometry yielded $J$=10.37$\pm$0.01\,mag, $H$=9.05$\pm$0.04\,mag, and $K$=7.00$\pm$0.03\,mag, where the errors reflect the uncertainty in the photometric calibration. 
%A series of short exposures ($\leq3$\,s) was obtained and during the $K$-band observations the telescope was additionally defocused to avoid saturation. Bias and flat-field frames were used for standard reduction of frames in IRAF. Given the relatively high infrared brightness of U\,Equ, a correction for nonlinearity was applied. The photometric calibration was performed with the standard star SAO070237 and was verified with 2MASS magnitudes of six field stars. Aperture photometry yielded $J$=10.55$\pm$0.03\,mag, $H$=9.16$\pm$0.06\,mag, and $K$=7.16$\pm$0.05\,mag. 
The $JHK$ systems used in earlier epochs may not be the same.

%------------------------------------------------------------------------------------------------
\section{Constraints from {\it Gaia} mission}\label{sec-gaia}
{\it Gaia} measured the U\,Equ's trigonometric parallax of 0.1891$\pm$0.0539\,mas and a proper motion of ($\delta_{\rm RA}, \delta_{\rm Dec}$)=(10.3512$\pm$0.0671, $-10.5706\pm0.0447$) mas\,yr$^{-1}$ \citep[all from DR3][]{gaiaDR3}. The parallax indicates a geometric distance of 4.8$^{+1.1}_{-0.9}$\,kpc and a photogeometric distance of 3.3$^{+1.2}_{-0.5}$\,kpc \citep{BJ}, more than 2--3 times larger than adopted by BOM96\footnote{For definitions of geometric and photogeometric distance determinations, see \citet{Bailer2021}}. We adopt here 4\,kpc as the distance to U\,Equ, as a compromise between the geometric and photogeometric method. Like for many other evolved and potentially binary stars, formal errors in both methods may be underestimated. 

Then, the Galactic latitude of U\,Equ implies that it is located 1.9\,kpc below the plane of the Galaxy and its projected motion is mainly away from the Galactic plane; the proper motion in the Galactic frame is ($\delta_l, \delta_b$)=(--3.4, --14.4) mas\,yr$^{-1}$. Adopting the radial velocity of --78\,\kms\ \citep{Geballe}, the Galactocentric motion\footnote{\url{https://docs.astropy.org/en/stable/coordinates/galactocentric.html#coordinates-galactocentric}} of ($\varv_x, \varv_y, \varv_z$)=(--65.0, 11.8, --282.5)\,\kms\ shows a relatively high speed away from the Galactic center. This makes U Equ very likely a near-halo object, and as such it may be also metal poor. Other halo objects have been identified near the position of U\,Equ \citep[e.g.][]{haloMap} and the Phlegethon stream is located nearby \citep{stream}. The interstellar extinction toward U\,Equ is, however, very low. Using the maps of \cite{extinction}\footnote{\url{https://irsa.ipac.caltech.edu/applications/DUST}}, we find the most likely value $E(B-V)$=0.078\,mag (or $A_V$=0.24 mag). BOM96 adopted a similarly low reddening of 0.09\,mag. A more recent 3D dust map of \cite{bayestar} provides a slightly higher value, $E(B-V)$=0.115\,mag (or $A_V$=0.36 mag).
%https://docs.astropy.org/en/stable/generated/examples/coordinates/plot_galactocentric-frame.html
%http://gala.adrian.pw/en/latest/coordinates/index.html
%https://dustmaps.readthedocs.io/en/latest/examples.html

In the 120 years between the Gaia observations and the first photographic plates, U\,Equ should have moved by about 1\farcs5. This figure is consistent with the appearance of the plates, but their seeing is of a similar order.

Gaia DR3 results for U\,Equ include a flux-calibrated spectrum in the range 3330--10200\,\AA\ at a low spectral resolution. In Sect.\,\ref{sec-salt-anal} and Fig.\,\ref{fig-allSEDs}, we make use of a combined BP/RP averaged spectrum.

%(The position relative to field stars looks similar in modern images and in the 100+ plates, so the motion is not huge.)

%----------------------------
\section{ALMA observations}\label{sect-alma}
For reference, we also extracted observations of the CO 3--2 pure rotational transition near 345.8\,GHz. It was found in the archive of the Atacama Large sub-Millimeter Array (ALMA). The data were obtained for the NESS collaboration \citep[PI P. Sciculna;][]{ness} on 10 Aug 2021 with the Atacama Compact Array (ACA) at an angular resolution of 3\farcs5. The data were processed with the CASA pipeline using default calibration tables. Next, data for spectral range covering the CO transition were imaged using tCLEAN with various visibility weighting schemes. In Fig.\,\ref{fig-masers1}, we present the source-averaged spectrum of CO 3--2. With a 302\,s exposure, the spectrum has a modest signal-to-noise ratio (S/N) and Fig.\,\ref{fig-masers1} shows a smoothed version.

The same line was observed by us with the Atacama Pathfinder Experiment 12 m submillimeter telescope (APEX) with the FLASH instrument \citep{flash} on 29 June 2012 and was detected at an even lower S/N of $\leq$10. Its intensity was close to that measured with ALMA (7.1$\pm$0.5\,Jy\,\kms), but uncertainties are large. This however indicates that there were no major changes in the CO feature since 2012.

The ALMA dataset shows a weak continuum source of 8.07$\pm$0.92\,mJy (1$\sigma$ error). The continuum was integrated within 330.2--333.7 and 343.2--346.2\,GHz avoiding the CO 3--2 emission. The continuum flux corresponds to an average wavelength of 885\,\um.
%% KMM: Is this continuum flux consistent with the long wavelength extension of the dust SED?
%%%%%%%%%%%%%%%%%%%%%%%%%%%%%%%%%%%%%%%%%%%%%%%%%%%%
\section{Effelsberg observations}\label{sect-sio}
\paragraph{SiO maser} U\,Equ was observed with the 100\,m Effelsberg radio telescope on 11 June 2023 using the S7mm Double Beam receiver covering 40--45 GHz. The data were acquired and calibrated using standard procedures. After 47 min of observing time on source, we detected a weak (0.24\,K $T_A$) and narrow (FWHM=1.4\,\kms) line of SiO $\varv$=1 $J$=1--0 at $V_{\rm LSR}$=$-75.0$\,\kms. Given the narrow line width, it is readily a maser line. It is  shown in Fig.\,\ref{fig-masers1}. Although other transitions (from SiO and other molecules) were covered in the same spectrum, for example SiO $\varv$=0 $J$=1--0, they were not detected at a nominal rms noise level of 25\,mK ($T_A$). 

\paragraph{H$_2$O masers}
The 100\,m telescope also observed the water transition $J_{Ka,Kc}$=$6_{1,6}$--$5_{2,3}$ at 22.2\,GHz (or 1.35\,cm) using the S14mm Double Beam receiver and the WFFTS backend. On 2 Nov. 2022, an integration of 19.5 min resulted in an rms noise of 28\,mK (in $T_A$) and displayed at least 3 emission lines centered at LSR velocities of $-88.2$, $-68.3$, and $-66.4$\,\kms; three other features are seen near $-93.6$, $-62.8$, and $-53.4$\, \kms\ at an S/N of 6--8. The stronger lines have FWHMs between 0.6--1.1\,\kms, indicative of masers. The strongest component has a peak intensity of 8.7\,K. The observations were repeated  on 14 Aug. 2023 with a total integration time of 29.4\,min and a resulting noise rms of 79\,mK. The spectrum of U\,Equ showed much stronger lines at this later epoch. The spectra are compared in Fig.\,\ref{fig-masers1}.  In Aug 2023, four strong components are seen at $-89.6$, $-87.8$, $-68.7$, and $-66.6$\,\kms. There is also a weaker line centered at --61.4\,\kms. The strongest component reaches 36.4\,K in the peak. The line widths are in the range 0.6--0.8\,\kms\ but the weakest feature appears broader with a FWHM of 1.9\,\kms. The strong variability in intensity and position of the H$_2$maser components is obvious between 2022 and 2023, but is even more striking when compared to the literature data (Sec.\,\ref{sec-masers}). Some of the early observations might not even cover the most redshifted components. Overall, the H$_2$O maser components appear over the velocity range spanning from about $-90.5$ to $-60.0$\,\kms, with the midpoint near --75\,\kms, which is also the position of the SiO maser line.

%%%%%%%%%%%%%%%%%%%%%%%%%%%%%%%%%%%%%%%%%%%%%%%%%%%%
\section{Photometric analysis}\label{sect-phot-analysis}
The collected archival data in Fig.\,\ref{fig-lc} show that U\,Equ is currently the brightest at visual and IR wavelengths since observations of this object started in 1904. Although it is not straightforward to compare the multiepoch data obtained in different photometric systems and without the knowledge of proper measurement uncertainties, it is apparent, especially in the visual bands, that the brightness of U\,Equ has been monotonically increasing since about 2009. The $JHK$ measurements probe a long period (1989--2006 and 2023), but unfortunately do not cover well the period of the gradual visual rise. The NOT $JHK$ measurements from 2023 are 1.1--1.5\,mag lower than in 1989, indicative of an increased near-IR flux by a factor of 2.8--4.0. The steady rise is also evident in the WISE and NEOWISE observations, but it is visibly steeper in the $W1$ band, that is  at shorter wavelengths. 

\begin{figure*}
    \centering \sidecaption
   \includegraphics[width=12cm]{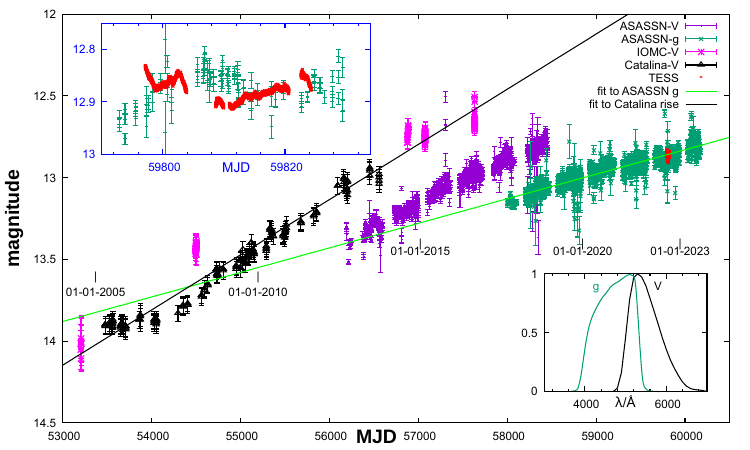}
    \caption{Rise in brightness of U\,Equ in visual bands. The blue inset shows variability documented by arbitrarily scaled TESS data. The inset in the lower right compares the transmission curves of the $V$ and $g$ filters.}
    \label{fig-rise}
\end{figure*}

The steady rise is best documented by combined observations obtained with different visual filters, albeit it is not covered entirely by consecutive observations in a single band. The visual rise is shown in more detail in Fig.\,\ref{fig-rise}. There, linear fits to the Catalina $V$ data after 2006 and to all ASASSN $g$-band data demonstrate that the rise had been steeper at its onset and has flattened at the time of the most recent observations. The total magnitude change in the visual is probably $\gtrsim$1\,mag. Although at a low photometric accuracy the rise appears to be relatively steady, the TESS data show small-amplitude flux variations of $\sim$5\% on a timescale of days. 

\subsection{SED evolution}
In an attempt to investigate the nature of the photometric changes, we selected three ``effective'' epochs with mid-IR data. Based on the data presented in Sec.\,\ref{sec-photometry}, we compiled the SEDs as follows.

\paragraph{Epoch 2023} The 2023 compilation is defined by the epoch of the NOT $JHK$ measurements on 25 Apr. 2023. We extrapolated the NEOWISE data to the same date by fitting linear functions to NEOWISE data only. We assign those points large uncertainties of 0.5 mag, which reflect the fit and the large scatter in the input data. We selected ASASSN $g$-band measurements from two dates closest to the date of NOT observations (22 and 29 May) and averaged them to obtain a representative optical-wavelength value. Finally, we also add to the SED the ALMA continuum measurement from 2021. Although it was obtained nearly two years earlier, the submm flux is not expected to have changed much. Even with a large uncertainty, its addition to the 2023 SED puts useful constraints on the models attempting to explain the source energetics in the most recent epoch.

\paragraph{Epoch 2010} For the defining moment of the 2010 SED, we select the epoch of WISE observations in all four bands. Nearly simultaneous with the WISE measurements (within a week), observations were made in the $yz$ bands as part of PAN-STARRS and in $V$ within the Catalina survey. We also add to this epoch $JHK$ fluxes obtained from linear interpolations between the latest measurements from the Arecibo survey and those from NOT. Again, we assign large uncertainties to these data points.

\begin{figure}
    \centering
%    \includegraphics[width=\columnwidth]{plotAllSEDs.pdf}
%    \caption{Comparison of SEDs at three epochs. For clarity, errorbars are not shown and data points are connected by lines. Connectors do not represent the actual shape of the SEDs. The ALMA measurement is not shown.}
    \includegraphics[width=\columnwidth]{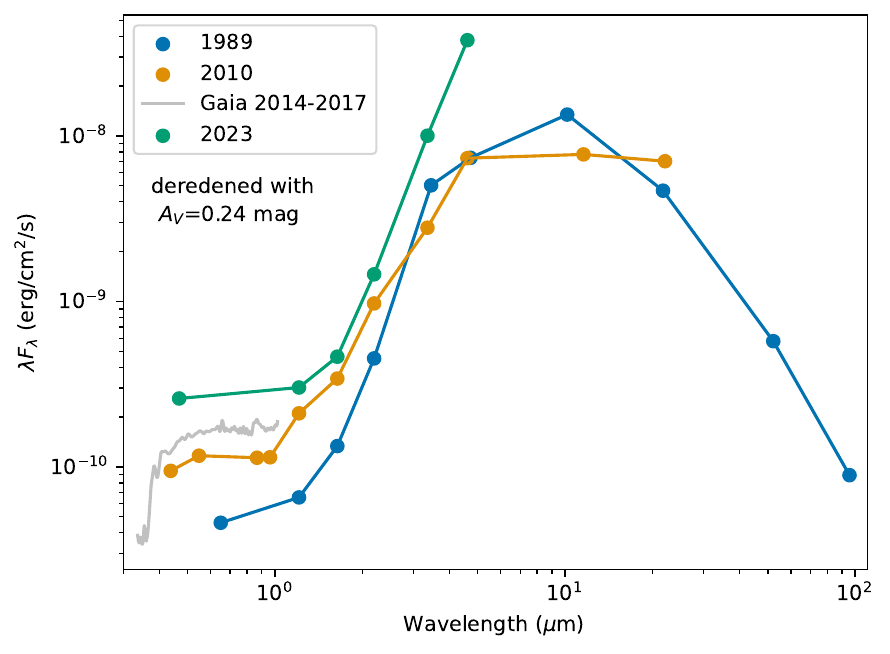}
    \caption{Comparison of SEDs at three epochs and to the Gaia average DR3 spectrum. For clarity, errorbars are omitted and data points are connected by lines. Connectors do not represent the actual shape of the SEDs. The ALMA measurement is omitted for display reasons. Data were corrected for interstellar extinction.}
    \label{fig-allSEDs}
\end{figure}

\paragraph{Epoch 1989} The earliest SED worth considering is defined by the simultaneous $JHKLM$ measurements obtained with the ESO 1\,m telescope in 1989. Although the IRAS flux measurements were obtained a few years prior, we add them to the 1989 epoch too, assuming no significant changes in the mid- and far-IR.had occurred over the relevant time period. We also include the magnitude from 1990 measured on the red POSS plate. The three SEDs are compared in Fig.\,\ref{fig-allSEDs}.

All data used in SEDs were corrected for the interstellar extinction, assuming $A_V$=0.24\,mag (see Sec.\,\ref{sec-gaia}). In order to understand the nature of the changes taking place in U\,Equ, we modeled these three SEDs using different dust radiative transfer tools. First, we attempted to reproduce the SEDs with models in which the central star is surrounded by a spherical shell of silicate dust. We used DUSTY \citep{dusty} model grids compiled by the Dusty Evolved 
Star Kit (DESK) \citep{desk}. Best fits were searched with the $\chi^2$ test. Although solutions were found, they were hardly satisfactory. The spherical models with different dust opacities always predict too strong fluxes at near-IR wavelengths, producing a spectral slope that is not observed in U\,Equ. From this, we conclude that spherical models are not adequate for reproducing the geometry of the U\,Equ system. Despite these caveats, for a distance of 4\,kpc, the models converge at luminosities of 8\,700 (epoch 2010) to 55\,000 (epoch 2023). Also, the estimated gas mass-loss rates increased from 10$^{-5}$ to 10$^{-4}$ M$_{\sun}$/yr between 2010 and 2023. Differences in the models likely reflect insufficient spectral coverage and inadequacy of the models, rather than a physical change in the system.

We next used a much wider grid of circumstellar SEDs compiled in \cite{robatille}. Although the database was created mainly for young stellar objects \citep{RobitailleArt}, it is applicable to a wider range of circumstellar environments, including evolved systems. A wide range of models exists in the grid with different sets of structural components, including a star, a flared passive disk with or without an inner hole, an envelope with or without cavities, and an ambient medium \citep[see][for a full overview]{RobitailleArt}. We performed analysis only for models with 4--9 free parameters. The models were compared to data points corrected for  interstellar extinction. Without any other constraints, the {\tt sedfitter} search was typically indicating best models with a giant star of effective temperature, $T_{\rm \star}$, from 10\,000 to 30\,000\,K (B spectral types) and an extra extinction of $A_V$=2 to 3\,mag. These synthetic SEDs reproduce the observations very well. However, since both $T_{\rm \star}$ and $A_V$ of these models disagree with spectroscopic results  (Sect.\,\ref{sec-salt-anal}) and with our constraints on the ISM extinction (Sects.\,\ref{sec-gaia} and \ref{sec-salt-reddening}), we discarded these models. 

We then narrowed the search for best Robatille SEDs by limiting the effective temperature to lower than 9000\,K and allowing for an extra interstellar reddening of $A_V \leq$0.1\,mag (cf. Sect.\,\ref{sec-gaia}). The best solutions are shown in Fig.\,\ref{fig-seds-sedfitter}. The fits for a wide range of circumstellar configurations are not reproducing the observations as well as the unconstrained {\tt sedfitter} results, but are still better than the spherical DUSTY models. For example, Table\,\ref{tab-sedfitter} presents parameters for models where the star is surrounded by a passive disk extending out from the sublimation radius (grid {\tt sp--s-i}) or by a transition-type disk with an inner hole (grid {\tt sp--h-i}). Although U\,Equ has been described in the past as a system possessing a disk, ours are the first results of actual SED fits encompassing a disk. Because none of the considered models satisfactorily reproduce the observations, we critically examine some of the model parameters.      

\begin{table*}[]
    \centering\caption{Best star-disk models reproducing the SEDs.}
    \label{tab-sedfitter}

    \begin{tabular}{l c c c c c c}\hline\hline
                               & \multicolumn{3}{c}{No hole models} & \multicolumn{3}{c}{Disk with inner hole} \\
                               & 1989 & 2010  & 2023  & 1989 & 2010   & 2023 \\
                               \hline
extra $A_V$ (mag)              &  0.100  & 0.030   & 0.022   &  0.100  &  0.100   & 0.097\\
Stellar radius (R$_{\sun}$)    &  81.3 & 36.8  &70.1   & 96.7  &  32.4  & 94.1\\
Stellar temperature (K)        & 7200  & 8100  & 5700 & 6700  &8200   & 6800\\
Luminosity at 4\,kpc (10$^3$\,L$_{\sun}$) & 33&  7.5& 19&  43&  5.9& 41\\
Disk dust mass (10$^{-3}$ M$_{\sun}$)  &2.7&12&1.0&0.48&32 &1.4\\	
Disk inner radius (AU)	       &$R_{\rm sub}$&$R_{\rm sub}$&$R_{\rm sub}$
                                                &3.0    & 37.2    & 1.3\\
Disk outer radius (AU)	       & 2922    &173   &599   &468      &601     &299 \\
Disk flaring power, $\beta$    &1.002   &1.070  & 1.118 & 1.119   &1.085   &1.033  \\
Disk surface density power     &--1.73  &--0.37 &--1.27 &--1.09   &--0.45  &--0.74  \\
Disk scaleheight (AU)          & 9.9    &18.8   & 18.3    & 19.1    &18.2    & 12.6  \\
Disk inclination (\degr)       & 71.2 & 63.5 & 59.1 & 61.6 & 64.3 & 71.3 \\
\hline\hline
    \end{tabular}
    \tablefoot{Disk in models with no inner hole starts at a respective dust sublimation radius, R$_{\rm sub}$. The luminosity was calculated from the radius and temperature using the Stefan-Boltzmann law without the Robatille's scaling factor \citep[cf.][]{RobitailleArt}.}
\end{table*}

\begin{itemize}
    \item Although some types of models give the best result for the highest allowed reddening of $A_V$=0.1\,mag, most of the best fits indicate a much lower extra extinction with an average of 0.05\,mag. Our assumed extinction of $A_V$=0.24\,mag is likely not far from the true value. 

    \item Across the different types of models, the best stellar temperature and radius are in the ranges of 8200$\pm$860\,K and 50$\pm$18\,R$_{\sun}$. From the Stefan-Boltzmann law, these indicate a luminosity of about 9800\,L$_{\sun}$. The uncertainties are too large to investigate how these stellar parameters were changing over time. 

    \item The models are not particularly constraining on the properties of the circumstellar matter. If a disk were the dominant component, Table\,\ref{tab-sedfitter} implies that: its mass is on the order of 10$^{-4}$ to 10$^{-2}$\,M$_{\sun}$; it extends from a few AU to a few hundred AU; and has a high inclination, that is, 59\degr--71\degr\ (so it is seen nearly edge-on).

\end{itemize}
The system can be characterized better only with more densely spaced contemporaneous observations, spanning from blue optical to submillimeter wavelengths.

% # # # # # # # # # # # # # #  SPECTRA # # # # # # # # # # # # # ## # # # # # ## # # # # # # #####
\section{Spectroscopy with SALT}\label{sect-salt-obs}
We used the High Resolution Spectrograph \citep[HRS;][]{HRS} at the Southern African Large Telescope (SALT) to observe U\,Equ on: 5, 12, 13 May 2002; 5, 8, and 9 Aug. 2022; 28 May and 30 June 2023. The typical exposure time was 1862 s. The HRS is a dual-beam \'echelle spectrograph operating in two arms that cover 370--550 and 550--890\,nm. We used the high resolution mode which afforded a resolution of about 65\,000. Because HRS is a fiber-fed instrument with a separate fiber observing the sky at a certain angular distance from the science target, subtraction of sky emission lines is not always perfect, which affected our spectra. The telluric lines are, however, easily recognizable when spectra from different epochs are compared at the heliocentric rest frame. Data were reduced and wavelength-calibrated using a HRS pipeline\footnote{\url{https://astronomers.salt.ac.za/software/hrs-pipeline/}} with default settings \citep{HRSpipeline}. \'Echelle orders and spectra from different arms were merged. Owing to the telescope design and mode of operations, SALT observations are not calibrated in flux. The pipeline-processed spectra were thus normalized to the continuum, that is, were divided by high-order polynomials. Spectra presented here were shifted to the heliocentric velocity frame. Normalized spectra from different dates were averaged. Due to a significant difference in the heliocentric velocity correction during different observing seasons, telluric features are severely broadened in the combined spectra. No reddening correction was applied.

%Using an averaged spectrum of U\,Equ from {\it Gaia} DR3, we attempted to flux calibrate the SALT spectra. The normalized spectra were smoothed to the spectral resolution of the Gaia data and divided by the flux-calibrated spectrum. Regions of strong telluric contamination were interpolated. Such a transfer function was next applied to the SALT spectra in full resolution. Since SALT and Gaia observations were not simultaneous, this should be treated as a very rough calibration. 
%----------------------------
\section{Analysis of the visual spectra}\label{sec-salt-anal}
\begin{figure}
    \centering
    \includegraphics[width=\columnwidth]{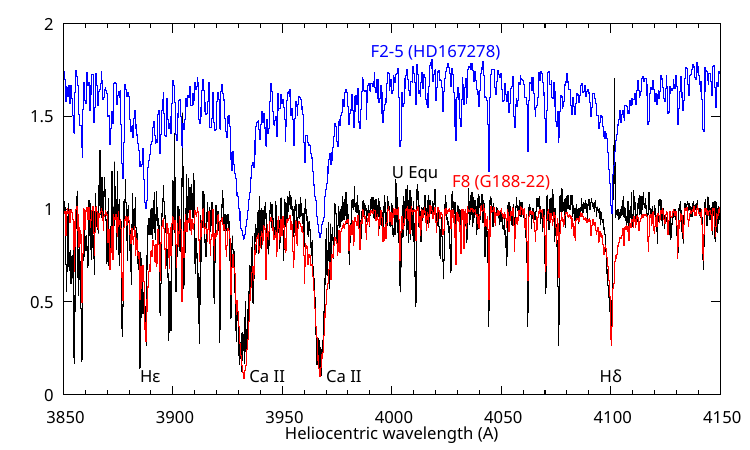}
    \caption{Comparison of spectra of the F-type stars HD167278 (blue) and G188$-2$22 (red) from XSL to the current spectrum of U\,Equ (black). U\,Equ is likely to have a spectral type F as well. }
    \label{fig-sptype}
\end{figure}

The full spectrum of U\,Equ averaged over all 2022--2023 exposures is shown in Figs.\,\ref{fig-atlas-p1}--\ref{fig-atlas-p4}. As mentioned, the spectrum has changed dramatically since its observations in the 
1990s. Instead of a plethora of molecular bands in emission and in absorption, we observe mainly circumstellar features of atomic gas of a low ionization. The atomic spectrum is very rich. We identified more than 600 features, most of which are absorption lines, but pure emission lines and composite P\,Cyg profiles are present as well. We identified the lines using several catalogs, chiefly from the NIST\footnote{\url{https://physics.nist.gov/PhysRefData/ASD/lines_form.html}}, which provides an interface for grouping lines in multiplets. A list of identified lines is available from the authors upon 
request, but the identification is also graphically presented in Appendix\,\ref{appendix-atlas}. Among the identified species are \ion{H}{I}, \ion{Fe}{i}, \ion{Fe}{ii}, \ion{Ti}{i}, \ion{Ti}{ii},  \ion{Cr}{I}, \ion{Cr}{II}, \ion{Sr}{ii}, \ion{V}{I}, \ion{Ca}{i}, \ion{Ca}{ii}, \ion{Ni}{I},  \ion{Mg}{I},  \ion{Si}{I}, \ion{Si}{II}, \ion{O}{I}, \ion{C}{I}, \ion{N}{I},  \ion{Ba}{II}, \ion{Co}{I}, \ion{K}{I}, and \ion{Na}{I}. We found no helium lines. Below, we highlight main groups of lines visible in the spectrum.
\begin{itemize}
    \item The Balmer and Paschen series of \ion{H}{I} are clearly recognizable in the spectra of U\,Equ. The lowest Balmer transitions, including H$\alpha$ and H$\beta$, have strong emission centered near a heliocentric velocity of --94\,\kms. Pure absorption is seen in lines from higher energy levels in both series. Based on the profiles, the absorption components are in most cases non-photospheric and form in an outflow crossing the line of sight. Additionally, H$\alpha$ has a complex emission-dominated profile with extended wings ($\approx$50\,\AA\ wide at the base) which are very likely formed due to photon scattering on electrons in the recombining plasma. Although H$\alpha$ is by far the strongest emission feature in the entire spectrum, it contributes only 9\% to the total flux in the $R$ band. The H$\beta$ line has a classic P\,Cyg profile, but photospheric absorption may be recognizable in the blue-shifted part.

    \item The only other lines showing broad emission wings comparable to those of H$\alpha$ are lines of the \ion{Ca}{II} IR triplet (rest wavelengths at 8498, 8542, and 8662\,\AA). These lines have complex profiles, with multiple overlapping absorption components. Additionally, each line of the triplet is contaminated by overlapping emission of the Paschen series.

    \item The H\&K lines of \ion{Ca}{II} near 3950\,\AA\ are dominated by broad absorption components. These are photospheric features somewhat contaminated in the line cores by circumstellar emission. By comparing \ion{H}{I} and \ion{Ca}{II} H\&K features to a grid of stellar spectra with known spectral types, we find that a type close to F6 best represents U\,Equ's photosphere. However, all F types and early G types of different luminosity classes match these noisy photospheric features comparably well. In Fig.\,\ref{fig-sptype}, spectra of U\,Equ  and two other F-type stars are compared. The stellar photospheric temperature is likely around 5700\,K, but the small number of photospheric features and a modest S/N make this number very uncertain. %No other clear photospheric features are seen. 

    \item Spectral features of CNO elements are of special interest. Neutral carbon is seen through the forbidden emission near 8725\,\AA. Other lines of \ion{C}{I} may be present but lack an unambiguous identification. No features of \ion{N}{I} were found. Neutral oxygen is traced with two [\ion{O}{I}] emission lines near 5577 (weak), 6300, and 6363\,\AA, and with three permitted multiplets in absorption. The triplets at around 7775 and 8446\,\AA\ have their upper energy levels near 11\,eV, while the weaker multiplet near 6155\,\AA\ is from relatively highly excited (13\,eV) states. These features are dominated by absorption and, since lines of a given multiplet blend, they form very distinct wide features, known to spectroscopists mainly from spectra of post-AGB stars. Future studies may attempt to derive relative abundances of the CNO elements in the circumstellar gas using these CNO tracers.

    \item By the number of individual transitions, the spectrum is dominated by lines of \ion{Fe}{i--ii}  ($\approx$270 lines) and of \ion{Ti}{i--ii} ($\approx$170 lines). (We find no forbidden lines of \ion{Fe}{ii}.) The presence of such rich spectra makes unique identification of individual lines rather difficult. The Fe and Ti spectra show primarily absorption at shorter wavelengths and pure emission in the red part of our spectra, but mixed, P\,Cyg, profiles are also observed. Chromium, \ion{Cr}{i--ii}, has a similar spectrum. An analysis of the \ion{Fe}{I} pure emission lines under the assumption of local thermodynamic equilibrium (LTE) presented in Appendix\,\ref{appendix-tom} suggests that the neutral gas is excited at a temperature of about 4200\,K. 

    \item Several lines from the ground level are observed. Among them, the \ion{Na}{i} doublet, studied in more detail below, is the most complex and displays several absorption components overlapping with weak emission. The \ion{K}{i} optical doublet is present as well, mainly in weak absorption, but is severely contaminated by telluric features. Two resonance lines of \ion{Ba}{II} at 4554 and 4934\,\AA\ are identified in absorption. Other clear resonance lines include absorption lines of \ion{Ti}{I}, \ion{Sr}{II}, and \ion{Fe}{I}. The resonance lines of \ion{Fe}{I} have double profiles.

\end{itemize}

% campare to post-AGB stars
Based on the described characteristics of the optical spectra, the circumstellar medium of U\,Equ has a low ionization degree, but is certainly more ionized than it appeared 28 years ago. Such a low-ionization circumstellar spectrum is not uncommon among post-AGB stars. In Appendix\,\ref{appendix-atlas}, we compare the 2022--2023 average spectrum of U\,Equ to that of a post-AGB star HD\,101584 \citep{olofsson2019}. The spectrum of HD\,101584 which we use was extracted from the Xshooter Spectral Library\footnote{\url{http://xsl.u-strasbg.fr/index.html}} \cite[XSL,][]{xsl}. The visual spectrum of HD\,101584 has been well described in the literature, for instance in \cite{Siva1999,HansObjectSpectrum} and in \cite{Kipper}.\footnote{Whether HD\,101584 is a post-RGB or post-AGB star is a matter of debate and depends on the object's true luminosity/distance, but the post-AGB status of this F0 supergiant has been recently reaffirmed by \cite{gaiaPostAGB}.} The spectra of U\,Equ and HD\,101584 are remarkably similar in terms of the presence of atomic features and their general profiles, that is, when a feature has a P\,Cyg profile in U\,Equ, it often has this type of profile in HD\,101584, too. Some differences exist, partially related to the better quality of the Xshooter spectrum. A few pure absorption and pure emission features can be found in HD\,101584, but not in U\,Equ. Some of the differences can be attributed to the difference in the relative abundances of the CNO elements, but a dedicated study would need to confirm this. There are other spectra of well known post-AGB stars that are very similar to what is currently observed in U\,Equ, including the spectrum of the protoplanetary nebula M1-92 (Minkowski's Footprint)  described in \cite{similarSpectra}, which has an F2 central star.

\subsection{Kinematic components}\label{sec-salt-kine}
%potential central components
As mentioned,  H$\alpha$, H$\beta$, and the \ion{Ca}{II} IR triplet contain an absorption component centered at a heliocentric velocity --94$\pm$3\,\kms\ and with a FWHM of about 26$\pm$5\,\kms. It can be seen in Fig.\,\ref{fig-profiles-Ha}. This component extends below the local continuum level in several of these lines, so it definitely represents an absorption component rather than a gap between two emission components. At the same velocity appears the red edge of some absorption lines, for instance, in the Paschen lines  (Fig.\,\ref{fig-profiles-Paschen}) and in the \ion{Na}{I} doublet (Fig.\,\ref{fig-profiles-Na}) --- but not in absorption lines of \ion{Si}{II} that have the blue edge at a few \kms\ higher velocity. Near --94\,\kms\ is also the cross-over point from emission to absorption for several P-Cygni profiles, for instance, in the profile of \ion{K}{I} (Fig.\,\ref{fig-profiles-Na}) (but not in the P-Cyg profiles of \ion{Fe}{II}); the same velocity is also exactly in between the two absorption components of some of the \ion{Fe}{I} lines, which display a double absorption feature (e.g., Fig.\,\ref{fig-profiles-Na}). The heliocentric velocity of --94\,\kms\ corresponds to an LSR velocity --81\,\kms, which is close to the velocity of the blue component of the 1612\,MHz OH maser observed in 1987 and to the red component of the H$_2$O maser observed in 1995 (see Sect.\,\ref{sec-masers}). Based on the inventory of optical lines where the component is seen, it may represent the coolest circumstellar gas on the line of sight toward the star. Hereafter, we call it component A.

%emission component
Pure emission lines, like those of [\ion{C}{I}] $\lambda$8727 and [\ion{O}{I}] $\lambda$6363, are centered at a heliocentric velocity of --87$\pm$2\,\kms, so are redshifted with respect to the absorption component described above by $\approx$7\,\kms. The center of the emission lines is also the crossover point between absorption and emission in the P-Cyg lines of \ion{Fe}{II} and \ion{Ti}{II} (Fig.\,\ref{fig-profiles-TiII}); it also sets the red edge of absorption in \ion{Si}{II} (Fig.\,\ref{fig-profiles-Paschen}). Converted to an LSR velocity of $-74$\,\kms, it matches very well the position of the red component of the OH masers and the position of the water masers observed in 1986 (see Sect.\,\ref{sec-masers}); it matches also to the velocity of the SiO maser discovered in U\,Equ in 2023 and the midpoint of the H$_2$O maser components detected with the Effelsberg telescope (Sect.\,\ref{sect-sio}). Hereafter, we call it component B. 

%which is stellar
The CO line observed by ALMA is centered exactly between components A and B.  It is uncertain which (if any) of these represents the systemic velocity of U\,Equ, but since mm-wave emission of CO is undoubtedly thermally excited and is not subject to circumstellar extinction, it may best represent the ``average'' systemic velocity. Unfortunately, the absorption lines that we interpret as photospheric are too contaminated and of a too low S/N to provide the stellar velocity. However, a cross-correlation of the average spectrum near the \ion{Ca}{II} H and K lines against a synthetic spectrum of a giant with an effective temperature of 6000\,K yields a heliocentric velocity of --96.8$\pm$24.8\,\kms. The large uncertainty makes this result of limited use, but the stellar velocity seems to be closer to component A. This is puzzling, as circumstellar SiO masers are usually considered to be very good indicators of the stellar velocity. Of course, there may be two or more dynamical centers of the system, for instance two stars, which could account for such a complex kinematic structure.

%outflow
Many of the lines of neutral and ionized species contain a broad absorption feature, which in some cases is combined with emission, forming classical P-Cyg profiles. This is undoubtedly a signature of an outflow. In lines of hydrogen of the Paschen series, \ion{Na}{I}, \ion{Fe}{II}, \ion{Si}{II}, the two \ion{O}{I} triplets -- the absorption component spans from approximately --85 down to nearly --300\,\kms\ (see Figs.\,\ref{fig-profiles-Ha}--\ref{fig-profiles-TiII}). The outflow has thus a maximum (projected) expansion velocity of 215\,\kms. This value is comparable to the escape velocity of a giant of about one solar mass and a radius of 5--10\,R$_\sun$. Not all absorption tracers reach such a high velocity. For instance, the main absorption component in the resonance line of \ion{Fe}{I} $\lambda$5110 extends only to an expansion velocity of 35\,\kms\ (Fig.\,\ref{fig-profiles-Na}). It would thus appear that the high-velocity outflow is mainly visible in lines requiring higher excitation. This is however not a strict rule, as the high-velocity component is readily seen in the doublet of \ion{Na}{I}, an element that is highly ionized at higher temperatures.

\subsection{Reddening}\label{sec-salt-reddening}
In both lines of the sodium $D_1D_2$ doublet, we find an extra absorption component centered at a heliocentric velocity of about --8\,\kms\ that has an FWHM of 13\,\kms\ (at an instrumental FWHM of about 7.5\,\kms, its intrinsic with is of 10.6\,\kms). This narrow component, although weak (with an equivalent width, EW, of 0.034\,\AA), is also recognizable in the \ion{K}{I} resonance lines (see Fig.\,\ref{fig-profiles-Na}), but not in other lines. It may represent a cool circumstellar or interstellar component. Given the Galactic position of U\,Equ, the low radial velocity of this component supports its interstellar origin. Indeed, an \ion{H}{I} spectrum at 21\,cm toward U\,Equ obtained by the Leiden-Argentina-Bonn survey \citep{LABsurvey} shows a single feature at the same velocity as the sodium lines. The depth of the component in both lines of \ion{Na}{I} is nearly the same, with EWs of 0.269 and 0.229\,\AA. Since the line ratio in the optically thin case should be 2.0, the observed lines must be saturated. The line of \ion{K}{I} is, however, optically thin. Using the $E(B-V)$ vs. EW calibration from \cite{MunariZwitter} for \ion{Na}{I} and \ion{K}{I}, we obtain $E(B-V)$ of 0.10$\pm$0.2\,mag or $A_V\approx$0.3\,mag. This extinction is slightly higher than derived from extinction maps in Sect.\,\ref{sec-gaia}. % but is certainly much lower than solutions suggested by our SED disk modelling in Sect.\ref{sect-phot-analysis}. This can be understood if the extra extinction suggested by SED models is circumstellar.  

We also attempted to constrain the total extinction by comparing the intensities of the Balmer lines. The Balmer lines are too complex to disentangle emission from absorption, and the entire normalized profiles were measured for the comparison. The corresponding continuum levels (of 2.08$\pm$0.05 and 2.34$\pm$0.06 in units of 10$^{-14}$\,erg cm$^{-2}$\,s$^{-1}$\,A$^{-1}$) were taken from flux-calibrated Gaia spectra near the two lines. The H$\alpha$ to H$\beta$ flux ratio we obtain, 3.0$\pm$0.1, is within the uncertainties consistent with the Case B ratio of those lines at temperatures between 5\,000--10\,000 K. It is thus consistent with none or minimal extinction between us and the recombining gas, consistent with our assumed $A_V$=0.24\,mag.

\subsection{Optical line variability}
In principle, our spectra allow us to test for line variability over timescales from days to about a year. Although spectra acquired a few days apart show some differences in line intensities relative to the local continuum, uncertainties in normalization to the continuum level and due to the modest S/N are large and do not allow us to draw firm conclusions on the variability. (Note that slit losses and other instrumental erratic effects are irrelevant for the comparison of continuum-normalized spectra.) Over the year covered by the observations, it would appear, however, that emission in the strong recombination lines of \ion{H}{I} and in the \ion{Ca}{II} triplet is getting weaker with time by a few tens of \%. Changes are observed in the high-velocity wings of the strongest absorption lines. The absorption at heliocentric velocities between about --300 to --100\,\kms\ is generally getting weaker. This is seen in lines of ionized species, for example in the lines of \ion{Fe}{I} near 5018, 6347, and 6371\,\AA, in the hydrogen lines of the Paschen series, and in both of the \ion{O}{I} triplets near 7770 and 8445\,\AA; a similar change is seen also in the high-velocity gas probed by the resonance absorption in the \ion{Na}{I} doublet. It would appear that the change is not related to excitation or ionization degree and could be interpreted as geometrical thinning of the outflow along the line of sight, for instance, due to expansion.

%-----------------------------------------D I S C U S S I O N----------------------------------------
\section{Discussion}\label{sect-discuss}
\subsection{Are the molecules gone? Not quite}
The main result of this study is the realization that U\,Equ has undergone a major transition and lost its unique molecular optical spectrum that drew attention to the object in the first place. The disappearance of molecules could be perhaps most naturally explained if intensified or hardened radiation of the central star (or stars)  completely dissociated the molecules in the circumstellar environment. However, our observations at longer wavelengths, obtained with ALMA and the Effelsberg telescope, seem to contradict this naive interpretation. 

The ALMA observations show that some cool molecular gas remains around U\,Equ. Estimating its mass would require information on the (distance and) gas temperature, which is not available at the moment. The ``survival'' of this cool molecular gas with the simultaneous disappearance of optical molecular bands may suggest that perhaps only the inner parts of the molecular envelope or disk were dissociated. The weak rotational emission in CO 3--2 shows also that the cool molecular outflow has a higher expansion velocity than assumed in the literature for the wind of U\,Equ (based mainly on OH maser spectra). The half of the full width of the CO line is of about 30\,\kms, consistent with normal winds of AGB stars and OH/IR objects. Our Effelsberg observation of the water maser lines show that the masers are still present tin the system, however they display most bizarre variability. Circumstellar H$_2$O masers are often linked to shocks \citep[e.g.,][]{Tafoya}, but are known to be present in systems with relatively hard spectrum, for instance, near high-mass protostars. 

The rather surprising discovery of an SiO maser ($\varv$=1) made with the Effelsberg telescope complicates the picture even further, because, for an evolved star, the masing action is expected to occur only a few stellar radii from the stellar surface, as typically observed near M-type giants and supergiants \citep[see, for example][]{Cotton2004,Reid2007}. Optical spectra definitely show a hotter photosphere in U\,Equ, of F type (Sect.\,\ref{sec-salt-anal}) or even hotter if the unconstrained SED models can be given credit (Sect.\,\ref{sect-phot-analysis}). Any imaginable drastic deactivation of the molecular emission, which is seen in the optical and at an excitation temperature of $\approx$500\,K, would also likely switch off the SiO maser. These paradoxes can partially be explained if U\,Equ is a binary and contains an M-type mass-losing giant that has at least one order of magnitude lower luminosity than the F type giant and thus does not show any photospheric features. Then, the observed spectral changes would be related to clearing off the surroundings of the hotter star, while leaving the envelope of the M giant unaffected. However, we find such an interpretation rather unlikely, as an M giant with a wind thick enough to produce an SiO maser (densities of 10$^9$\,cm$^{-3}$) would also be luminous and should show some spectral features in the optical SALT spectra. 

Alternatively, SiO may arise in a disk or disk-wind, like in the peculiar young stellar object Source\,I in the Orion Kleinmann Low nebula (KL) \citep{Matthews2010,SourceI}. Therein, the central source is thought to be a progenitor of a B-type star \citep{Goddi11,adam} and may be characterized by a radiation field that is harsher than that of M-type giants and closer to that of the main stellar component of U\,Equ. The SiO masers are excited in an expanding and rotating disk-wind region of a size of 40\,AU with a possible involvement of circumstellar shocks. Because Source\,I belongs to a disintegrated multiple system with its former members running away from a place of a major eruption that took place $\approx$500 years ago, it was proposed that Source\,I underwent a major dynamical interaction at this time \citep{Gomez2008}. Some authors go as far as to propose that Source\,I is a product of a merger \citep{farias}. SiO masers like those in Source\,I are however very rare \citep{Zapata2009}. Can U\,Equ be in some aspect similar to this rare occurrence of circumstellar SiO masers? Has the transition in U\,Equ triggered the maser emission only recently? What is the relation of the SiO maser to the OH and H$_2$O masers observed decades ago in U\,Equ? Although a lot is unclear, it is apparent that U\,Equ is important for our understanding of maser evolution in post-main sequence systems.

With U\,Equ having lost its molecular emission bands that sparked interest in this object in the 1990s, there remain only a handful of objects with this type of spectrum. \cite{lloyd} discussed this small, but diverse group of objects, which includes the RV Tauri stars and the red supergiant VY\,CMa. However, in the 26 years since that discussion, the list of objects has grown somewhat longer. VY\,CMa remains the poster child of this phenomenon, with variable but %unthreatened
persistent emission bands of TiO, AlO, ScO, and VO \citep{Wallerstein2001,KamiAlO}. The Galactic red-nova remnants, V838\,Mon, V4332\,Sgr, V1309\,Sco, products of stellar merger events,  show similar emission bands \citep[e.g.,][]{KamiV4332}, but those are fading on time scales of decades. A few young stellar objects display emission bands after outbursts \citep{Herczeg,Hillenbrand}, but they are usually weak and fade away relatively fast. Mira stars also occasionally show transient emission bands from material shocked by pulsations \citep[e.g.,][]{KamiMiraAlO}. Molecular emission bands are therefore a very rare and -- for most objects -- a short-lived transient phenomenon. This was likely the case of U\,Equ.

%---------------------------------------------------------------------
\subsection{Nature of the photometric rise}

The remarkable revamping of the optical spectrum associated with the seemingly modest $\sim$1\,mag change in the optical and near-IR fluxes could have been caused by several phenomena. It appears to be possible that the change was simply caused by a geometric effect, namely, a reorganization of the circumstellar medium on the line of sight due to disk rotation or inhomogeneous and anisotropic mass loss. In some scenarios, such a reconfiguration would change the extinction and optical spectrum without any major changes in the physical properties of the star. However, the simultaneous rise at visual and mid-IR bands that we observe (with the caveat that the behavior at longer wavelengths is unknown) could suggest an increase in the object's bolometric luminosity. Therefore, we rather favor a scenario in which the transition was more fundamental than barely a change in circumstellar arrangement along the line of sight. We find it more likely that it is related to a major change in the circumstellar medium and to an increase of the total luminosity of the system due to physical changes in the star or due to activation of an accretion episode (on the main star or on its hypothetical companion).

The amplitudes of the rise in photometric brightness in the visual and near-IR bands that we find in U\,Equ are higher than amplitudes of typical long-term flux changes in other post-AGB stars. Photometric monitoring projects targeting several post-AGB stars show typical optical variations of $\lesssim$0.5 mag \citep[e.g.,][]{mesler,Arkhipova,Hrivnak2010,Hrivnak}. There are reports of changes that reach a higher amplitude, but these are almost always short-lasting drops in brightness with unremarkable changes in the optical spectrum. They can often be ascribed to eclipses, trivial episodes of obscuration (for instance, by dusty wind material), or to pulsations \citep[cf.][]{AGBvar,Hrivnak}. Such cases do not require any activation of an extra energy source. The case of U\,Equ is different because it has shown a continued, monotonic, long-lasting rise by 1 mag associated with a dramatic change in the visual spectrum. Such a high-amplitude monotonically rising brightness is certainly not common among post-AGB objects, as it requires an extra energy sources. Even if the relative change in the magnitudes scale is not huge, like here $\approx$ 1, it implies a flux increase by a factor of a few, which at the luminosity of $10^4$\,L$_{\sun}$ implies a considerable amount of extra energy.

The possibility that we caught U\,Equ's central star at the onset of the fast transformation from an AGB object to a proto-WD is most fascinating. However, with current data, we have no solid evidence of an increasing temperature at a nearly constant luminosity of the star. The quality and coverage of the available data is insufficient. However, we calculate that a black-body of a fixed luminosity would increase its optical flux near 5000\,\AA\ with an increasing temperature, so the observed photometric changes we observe are not inconsistent with an early transformation on the post-AGB. (For instance, a change from 4000 to 8000\,K increases the flux by a factor of 2.3.)

The best observed fast transition in post-AGB stars is a manifestation of a late He flash, where a cooling central star of a planetary nebula turns back to a configuration with temperature and luminosity characteristic of a post-AGB object. Several such `born-again' (a.k.a. late-thermal-pulse) objects are known (e.g., V4334\,Sgr, V605\,Aql, and FG\,Sge), and they all are quickly evolving carbon stars \citep[e.g.,][]{vanWinckelReview}. U\,Equ is a descendant of an O-rich OH/IR star (thus likely more massive than carbon stars) and the characteristics of its photometric changes cannot be ascribed to a late He-shell flash. A thermal pulse in an evolved massive AGB is another scenario that can be considered for U\,Equ but is hard to verify with the data currently in hand.  

Like in many other post-AGB stars, we may be witnessing in U\,Equ an activation of an accretion episode that increased the luminosity of the system, possibly triggered by binarity and matter being accreted from a circumbinary disk. Such systems are often associated with fast jets \citep{bollen}. There is no direct evidence of a binary companion to U\,Equ, but there hardly ever is for post-AGB stars with dwarf companions of a much lower luminosity. Any observational indication of jets in U\,Equ would confirm active accretion episodes, and thus can explain some of the changes seen in this object over the last decades. The short-term low-amplitude light variations observed in U\,Equ by TESS (Sect.\,\ref{sec-photometry}) is similar to flickering observed in accretion-active sources \citep{Scarangi}, supporting our hypothesis of activated accretion in U\,Equ.

%(There have also claims of long-lasting fading of post-AGB stars (russion paper), but these have never been confirmed with precise photometry...?) 

%Circumstellar molecular disks, like the one seen in the 1990s in U\,Equ, were long considered long-lived reservoirs of matter in post-AGB stars \citep{Omont2001}.  The current understanding of U\,Equ and V Car may change that assumption. Those disks may not be long-lasting. 

%The described case of U\,Equ can be also understood as a transition from late AGB phase with a high mass-loss, highest so far in the stellar lifetime, into post-AGB phase with a mass loss-rate lower by orders of magnitude. In fact, some definitions of the post-AGB phase require an apparent visual counterpart (Kwok),
\begin{figure*}
    \centering\sidecaption
   \includegraphics[width=12cm]{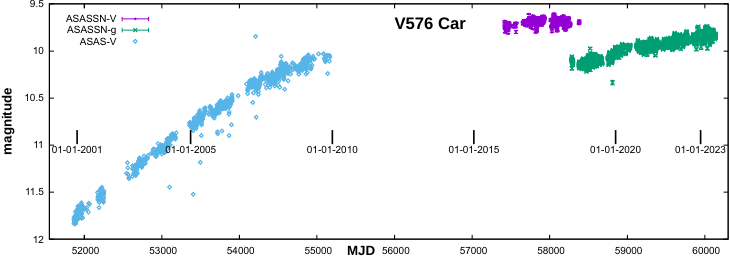}
    \caption{Visual light curve of V576\,Car extracted from ASAS and ASASSN data. The rise in brightness has continued since at least 1989 \citep{couch}.}
    \label{fig-lc-v576}
\end{figure*}

%--------------------------------------------------------------------------------------
\subsection{U\,Equ as a post-AGB object}
The general observational characteristics of U\,Equ, primarily its SED and its spectral type, make it most similar to the class of post-AGB and post-RGB objects known to harbor dusty disks. This class of objects has been  recently reviewed by \cite{KluskaReview}. More specifically, U\,Equ resembles their Category\,0 sources, which are associated with disks seen at high inclinations. If confirmed as a member, U\,Equ would be the first halo object in this group. The post-AGB systems discussed by \cite{KluskaReview} show a chemical anomaly whereby refractory metals are underabundant in the stellar atmosphere due to selective accretion from a circumbinary disk. Future studies of U\,Equ should investigate whether it displays a similar chemical peculiarity. %The similarity makes U\,Equ a very likely binary with and F--G type primary star. %Until then, the object is a candidate member of these class of objects. Whether the spectral and photometric transition observed over the last decades In U\,Equ is a common phenomenon for these objects, is an open question.

%These spectra are produced by interaction of the stellar (and accretion, if present) fluxes with the circumstellar medium. Since some of the mentioned post-AGB stars with spectra similar to U\,Equ are of spectral types F--G, one can suspect the same for the central star of the U\,Equ system. 

As noted in Sect.\,\ref{sec-salt-anal}, we find a particularity strong resemblance of U\,Equ's contemporary spectra to those of the post-common-envelope binary system HD\,101584. Multiple lines in the spectrum of HD\,101584 are variable \citep{Kipper} at a level similar to the changes observed in our SALT spectra. Spectral features arising in the circumstellar medium of HD\,101584 have multiple components, with a maximum projected expansion velocity of 130\,\kms. This velocity is only slightly lower than the extreme components of U\,Equ (215\,\kms) and comparable to the outflow velocity measured for many strong absorption lines (it is very likely that we look at the two systems from different angles and projection effects are important). The central stars of both systems may also be similar. Early UV spectra of HD\,101584 seemed to suggest a presence of a hot B9\,II star \citep{Bakker}, but later studies, based mainly on visual data, preferred a cooler star, of spectral types A--F \citep{Kipper}, again in accord to what we know about U\,Equ. HD\,101584, just like U\,Equ, has OH 1667 MHz masers.\footnote{The most recent detection observations are from \cite{lintel}.} They are located in a bipolar structure and display exceptionally broad line profiles \citep{lintel}. The projected OH expansion velocity of 42\,\kms\ is much smaller than the velocity of the CO outflow traced at (sub)mm wavelengths \citep{olofsson2019} and the outflow probed by optical lines. These unique masers bear strong resemblance to the unusual masers of U\,Equ (except for the spatial distribution, which for U\,Equ is unknown; see Sect.\,\ref{sec-masers}). It is unclear whether HD\,101584 has ever gone through spectral and photometric transformations such as those reported in this paper for U\,Equ. Since HD\,101584 has been considered to be a rare example of a system that has recently gone through a common-envelope phase \citep{olofsson2019}, we should consider U\,Equ as another unique system of similarly interesting recent evolution and therefore worth of further studies. 

%While widely assumed to be a post-AGB star, at the Gaia-derived distance of 1\,kpc, uncertain for a binary, the luminosity of the objects rather indicates a post-RGB status. 

%HD\,101584 is also considered to be a binary \citep{Kluska} with an orbital period of 218\,d \citep{Bakker}. While there is as yet no direct evidence of binarity in U\,Equ, just by the remarkable similarity to HD\,101584 it may be a binary as well. Also, HD101594 appears to be enhanced in C and N elements. Whether this is also the case for U\,Equ should be investigated in the future. It has a complex disk which is seen almost face on \cite{Kluska}.

%-----------------------------------------------------------------------------------------------------------

\subsection{V576 Car}\label{V576Car}
The closest known analog to U\,Equ, as it was known to 1990s observers, is V576\,Car (=IRAS\,08182--6000). Its optical spectrum, dominated by molecular absorption bands, but devoid of emission features seen earlier in U\,Equ was described in detail in \cite{couch}. The circumstellar absorption bands of all the main electronic systems of TiO, AlO, and VO were seen in the spectra of V576\,Car. The underlying stellar photosphere was not easy to identify (as in U\,Equ), but the best efforts indicated an F--G supergiant, similar to what we find for U\,Equ (Sect.\,\ref{sec-salt-anal}). Additionally, \cite{couch} reported that the $J$-band fluxes were monotonically increasing over the 1990s, totaling to a 2.5 mag amplitude over 10 years. We found visual photometric data displaying a continuation of this rise in brightness over the last two decades by another 2.5 mag. The data taken from the All Sky Automated Survey \citep[ASAS;][]{asas} and ASAS-SN sky patrol are presented in Fig.\,\ref{fig-lc-v576}. The steady rise in brightness seen in V576\,Car is reminiscent of the light curve of U\,Equ, although larger in amplitude. Additionally, TESS data for V576\,Car (not shown) display the same small-amplitude erratic variability as that we found in U\,Equ (Fig.\,\ref{fig-rise}). Most importantly, when we obtained a spectrum of this object on 16 May 2023 with HRS-SALT, the absorption bands of molecules had completely disappeared  and been replaced by a spectrum remarkably similar to the current spectrum of U\,Equ. The contemporary spectra of both objects are shown along each other in Figs.\,\ref{fig-atlas-p1}--\ref{fig-atlas-p4}. With little doubt, the changes in both stars are very similar and in both cases resulted in an object with a spectrum that is now more typical for post-AGB stars. The question that arises is how common is the behavior when a post-AGB star switches from a spectrum dominated by molecular bands into the atomic spectrum. 
%Is this just an episode related to a short-term reconfiguration of the circumstellar architecture, or is it a fundamental change directly related to the fast evolution of the central giant. Finally, can it be related to the long sought accretion events when gas is drawn from the circumbinary disk into the post-AGB component, adding to the "depletion" effect? Extra energy from accretion would naturally explain the steady rise in brightness of U\,Equ and V576\,Car. 
The absence of masers around V576\,Car \citep{couch} needs to be revisited with new sensitive observations. 

A larger sample of stars exhibiting the U\,Equ phenomenon would definitely help to understand better the nature of the changes. \cite{arfon} searched the southern sky for objects with similar characteristics to U\,Equ and V576\,Car (i.e., as it appeared in the 1990s) but found none that would exhibit circumstellar molecular features. The phase in which  U\,Equ and V576\,Car were caught must therefore be short-lasting, as expected for fast-evolving dust-embedded post-AGB stars.  Perhaps future searches can focus on the steady rise in brightness over timescales of $\gtrsim$3 decades as the identifying feature of the U\,Equ phenomenon. %There have been reports of such long-term variations in eveolved stars \citep[][]{Ganhi}, but have attracted little attension so far. 

\subsection{Summary and outlook}
More observations are needed to explain the changes in U\,Equ and V576\,Car. Continued photometric and spectroscopic monitoring can reveal how long and to what end the noticed changes will occur. Multiwavelength observations and interferometric imaging would be particularly useful for understanding the current configuration of the star or binary with respect to the disk and other circumstellar features. We should be also looking for more objects experiencing the U\,Equ phenomenon, as such studies  have the potential to expose the genuine and long-sought transition from AGB or OH/IR phase into a planetary nebula, or they can deepen our understanding of common-envelope interactions. 

\begin{acknowledgements}
T.K. acknowledges funding from grant SONATA BIS no.
2018/30/E/ST9/00398 from the Polish National Science Center. K.I. was supported by Polish National Science Center grant Sonatina 2021/40/C/ST9/00186.

Based on observations collected at the European Organization for Astronomical Research in the Southern Hemisphere under ESO program 266.D-5655(A). 

This work made use of the HDAP which was produced at Landessternwarte Heidelberg–K\"onigstuhl under grant no. 00.071.2005 of the Klaus-Tschira-Foundation. 

Some observations reported in this paper were obtained with the Southern African Large Telescope (SALT). Polish participation in SALT is funded by grant no. MNiSWDIR/WK/2016/07. 

Based on data from the OMC Archive at CAB (INTA-CSIC), pre-processed by ISDC and further processed by the OMC Team at CAB. The OMC Archive is part of the Spanish Virtual Observatory project. Both are funded by MCIN/AEI/10.13039/501100011033 through grants PID2020-112949GB-I00 and PID2019-107061GB-C61, respectively.

This research has made use of the SIMBAD database, operated at CDS, Strasbourg, France. 

This research made use of hips2fits, https://alasky.u-strasbg.fr/hips-image-services/hips2fits a service provided by CDS. 

The CSS survey is funded by the National Aeronautics and Space Administration under Grant No. NNG05GF22G issued through the Science Mission Directorate Near-Earth Objects Observations Program. The CRTS survey is supported by the U.S. National Science Foundation under grants AST-0909182 and AST-1313422.

The Pan-STARRS1 Surveys (PS1) and the PS1 public science archive have been made possible through contributions by the Institute for Astronomy, the University of Hawaii, the Pan-STARRS Project Office, the Max-Planck Society and its participating institutes, the Max Planck Institute for Astronomy, Heidelberg and the Max Planck Institute for Extraterrestrial Physics, Garching, The Johns Hopkins University, Durham University, the University of Edinburgh, the Queen’s University Belfast, the Harvard-Smithsonian Center for Astrophysics, the Las Cumbres Observatory Global Telescope Network Incorporated, the National Central University of Taiwan, the Space Telescope Science Institute, the National Aeronautics and Space Administration under Grant No. NNX08AR22G issued through the Planetary Science Division of the NASA Science Mission Directorate, the National Science Foundation Grant No. AST-1238877, the University of Maryland, Eotvos Lorand University (ELTE), the Los Alamos National Laboratory, and the Gordon and Betty Moore Foundation. 

This research has made use of the VizieR catalogue access tool, CDS, Strasbourg, France (DOI : 10.26093/cds/vizier). The original description of the VizieR service was published in 2000, AAS 143, 23.

Based on observations made with the Nordic Optical Telescope, owned in collaboration by the University of Turku and Aarhus University, and operated jointly by Aarhus University, the University of Turku and the University of Oslo, representing Denmark, Finland and Norway, the University of Iceland and Stockholm University at the Observatorio del Roque de los Muchachos, La Palma, Spain, of the Instituto de Astrofisica de Canarias

This research has made use of the Spanish Virtual Observatory (https://svo.cab.inta-csic.es) project funded by MCIN/AEI/10.13039/501100011033/ through grant PID2020-112949GB-I00.

\end{acknowledgements}
\bibliographystyle{aa}
\bibliography{0bib.bib}

\begin{appendix}
\section{Photographic plates}
\begin{figure*}
    \centering
   \includegraphics[width=\textwidth]{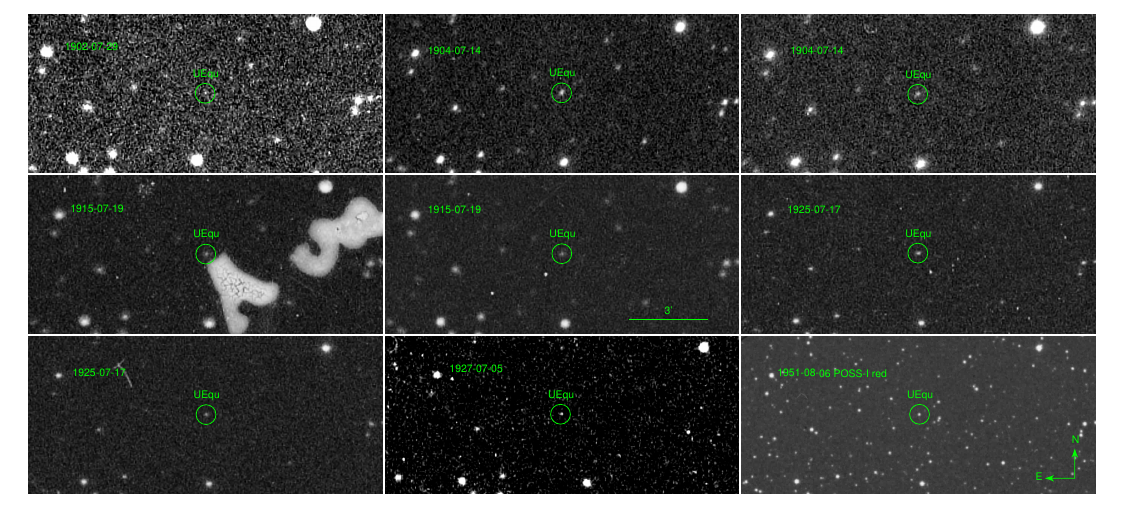}
    \caption{Sample Heidelberg-K\"onigstuhl blue plates. The cutouts are centered near U\,Equ (encircled). The dates of observations are given. For reference, a deeper POSS image is shown in the lower right corner. Images are displayed in a linear intensity scale.}
    \label{fig-plates}
\end{figure*}

The field around U\,Equ was captured by photographic plates throughout the first half of the 20th century. A sample of them is shown in Fig.\,\ref{fig-plates}.

\section{Spectral energy distribution models}
\begin{figure}
    \centering
    \includegraphics[width=\columnwidth]{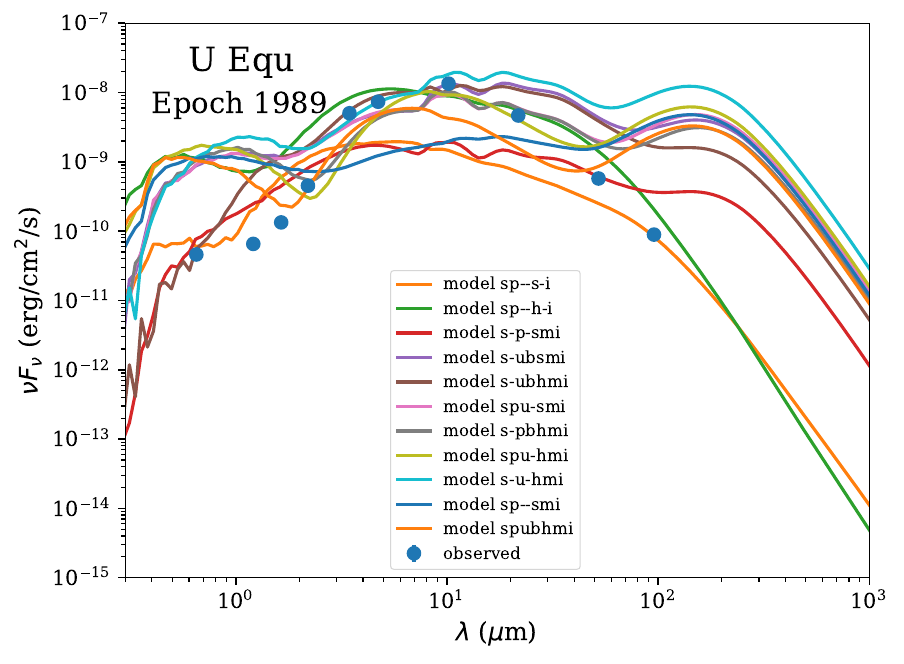}
    \includegraphics[width=\columnwidth]{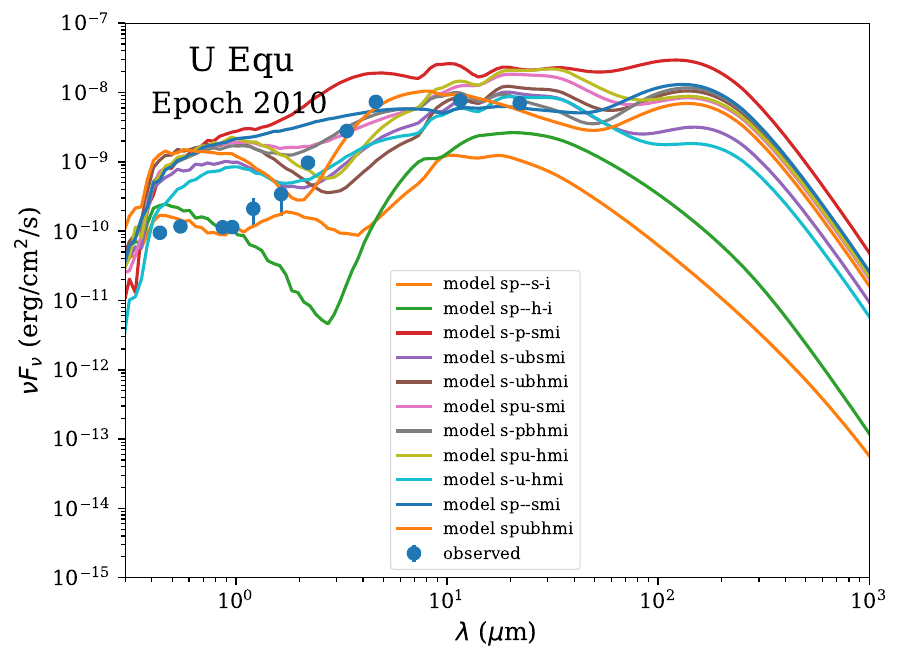}
    \includegraphics[width=\columnwidth]{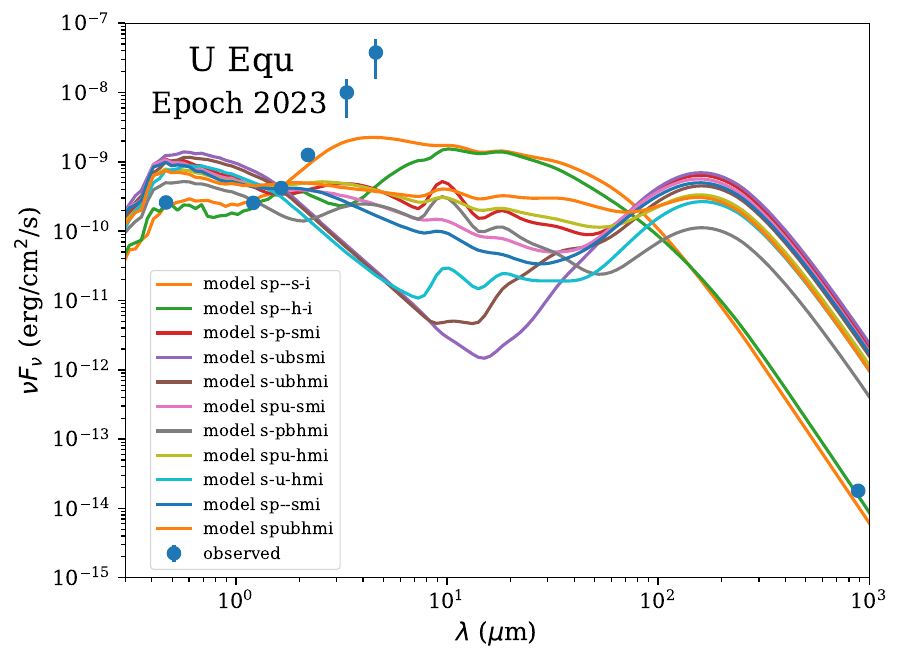}
    \caption{Model and observed SEDs of U\,Equ in 1989 (top), 2010 (middle), and 2023 (bottom). Observations are marked with points. Solid lines show best-fit SED models from {\tt sedfitter} for different circumstellar configurations descibed in \citep{RobitailleArt}.}
    \label{fig-seds-sedfitter}
\end{figure}

\section{Spectral atlas}\label{appendix-atlas}
Here we present an atlas of spectra of U\,Equ and V576\,Car compared to that of HD\,101584.
\begin{figure*}
    \centering
    \includegraphics[page=1, width=\textwidth]{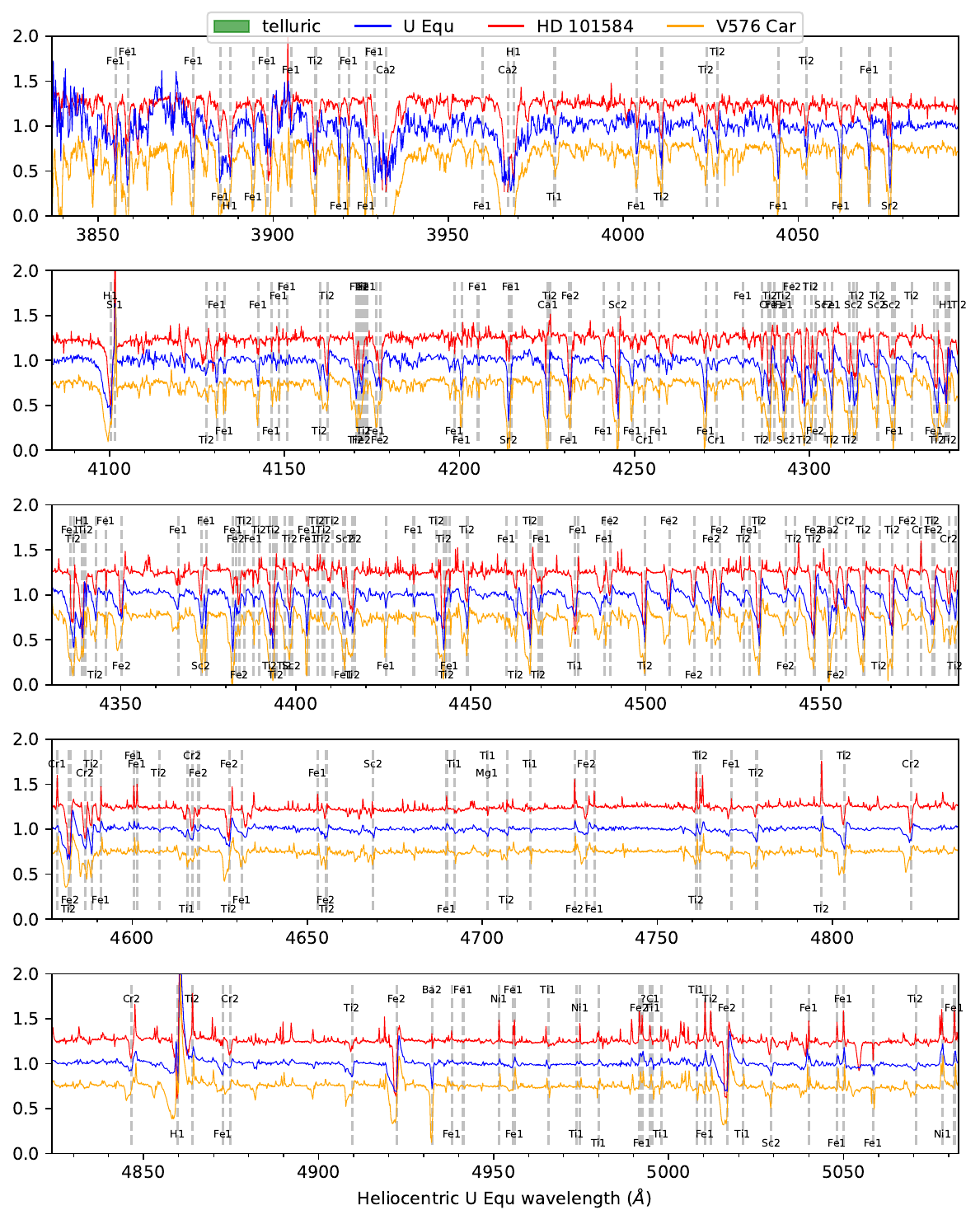}
    \caption{Average normalized spectrum of U\,Equ (blue) is compared to that of HD\,101584 and of V576\,Car (orange). Main telluric bands are marked in green. The vertical dashed lines and corresponding labels identify the main spectral features ("1" corresponds to neutral atoms and "2" to first ionization state).}
    \label{fig-atlas-p1}
\end{figure*}

\begin{figure*}
    \centering
    \includegraphics[page=2, width=\textwidth]{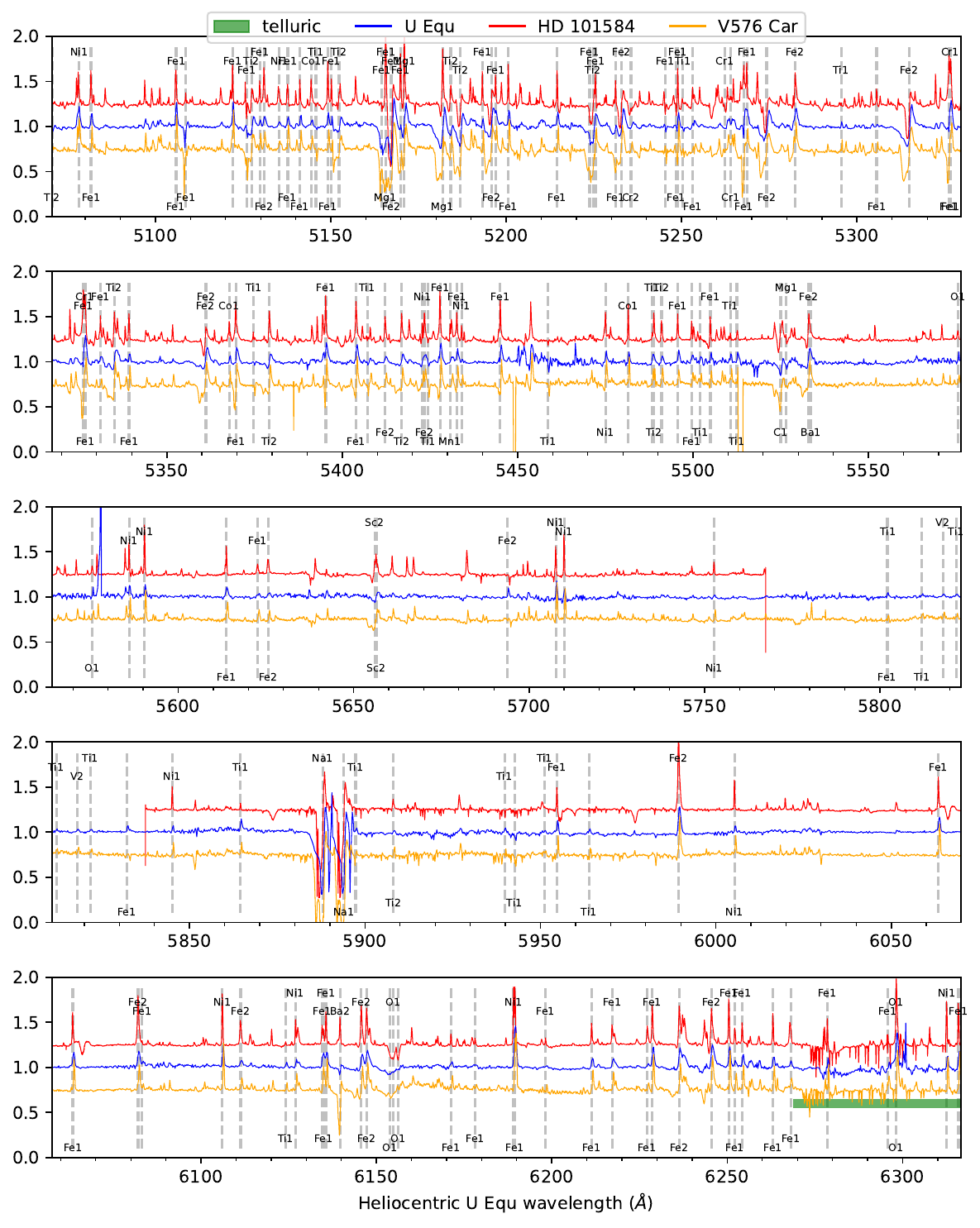}
    \caption{Continued.}
    \label{fig-atlas-p2}
\end{figure*}
\begin{figure*}
    \centering
    \includegraphics[page=3, width=\textwidth]{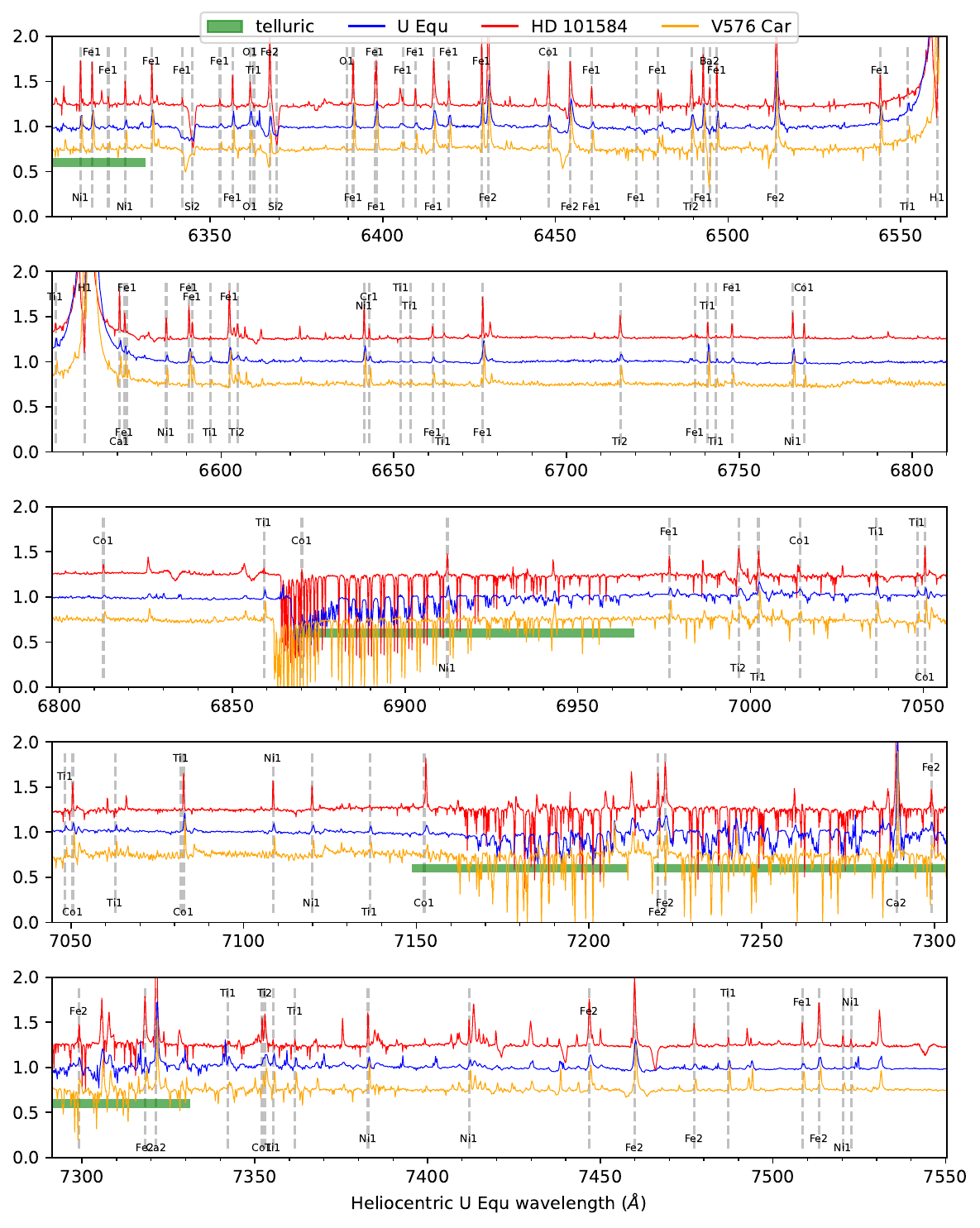}
    \caption{Continued.}
    \label{fig-atlas-p3}
\end{figure*}\begin{figure*}
    \centering
    \includegraphics[page=4, width=\textwidth]{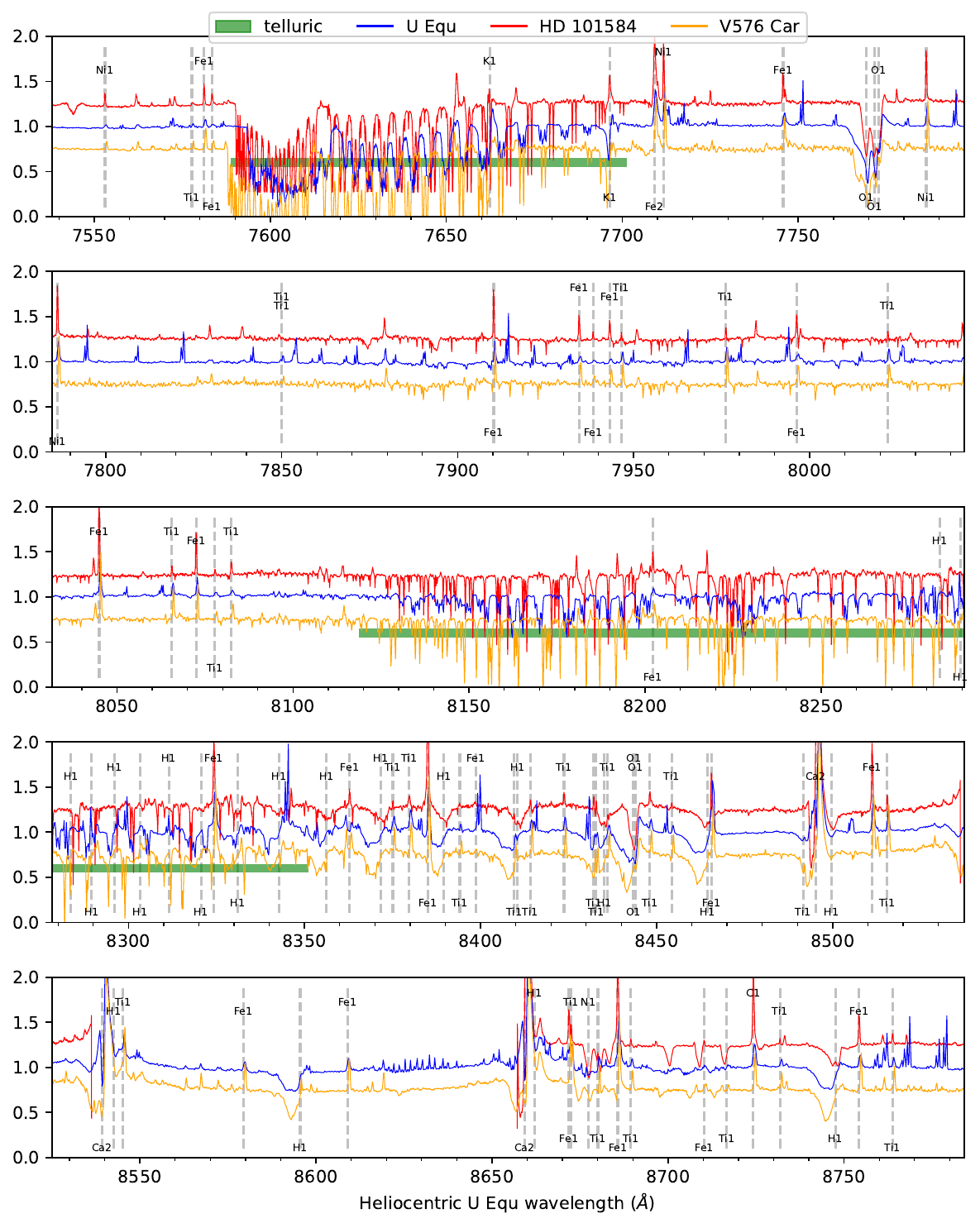}
    \caption{Continued.}
    \label{fig-atlas-p4}
\end{figure*}

\section{LTE analysis of \ion{Fe}{I} lines}\label{appendix-tom}
Using the "Saha-LTE Spectrum" tool available from the NIST online database, we simulated the intensities of \ion{Fe}{I} lines at different temperatures between 4000--10\,000\,K with a step of 200\,K. As the calculated line intensities are measured in arbitrary units, the intensities were scaled to the relative intensity of the line near 8514\,\AA, which has the highest S/N and is a blend of two strong transitions. We analyzed only lines in pure emission, with no blending or contamination from telluric or other features. The observed line intensities were measured in IRAF. A $\chi^2$ test was run to find the best fitting line ratios. The test yielded a temperature of 4200$\pm$200\,K.

As a test of opacity, we also compared the line ratios of lines in the same multiplet with the ratios of the transition probabilities, $A_{ki}$. If the \ion{Fe}{I} emission is mostly optically thin, the scaled fluxes and $A_{ki}$ values should be comparable. The resulting comparison shows some agreement of the NIST models within 5$\sigma$ errors, but not for all line pairs. The goodness of fit shows no trend with upper energy level, but the majority of lines used for this test originate from relatively close energy levels in range of 2--6\,eV. Out of the nine \ion{Fe}{I} ratios measured, four fit well within the 5$\sigma$ errors whilst three more do not lie much further from the observed values. Only two line ratios vary significantly (6230/7748 and 6494/9593). We conclude that the \ion{Fe}{I} emission is not very optically thick.

\section{Line profiles}

\begin{figure}
    \centering
    \includegraphics[width=\columnwidth]{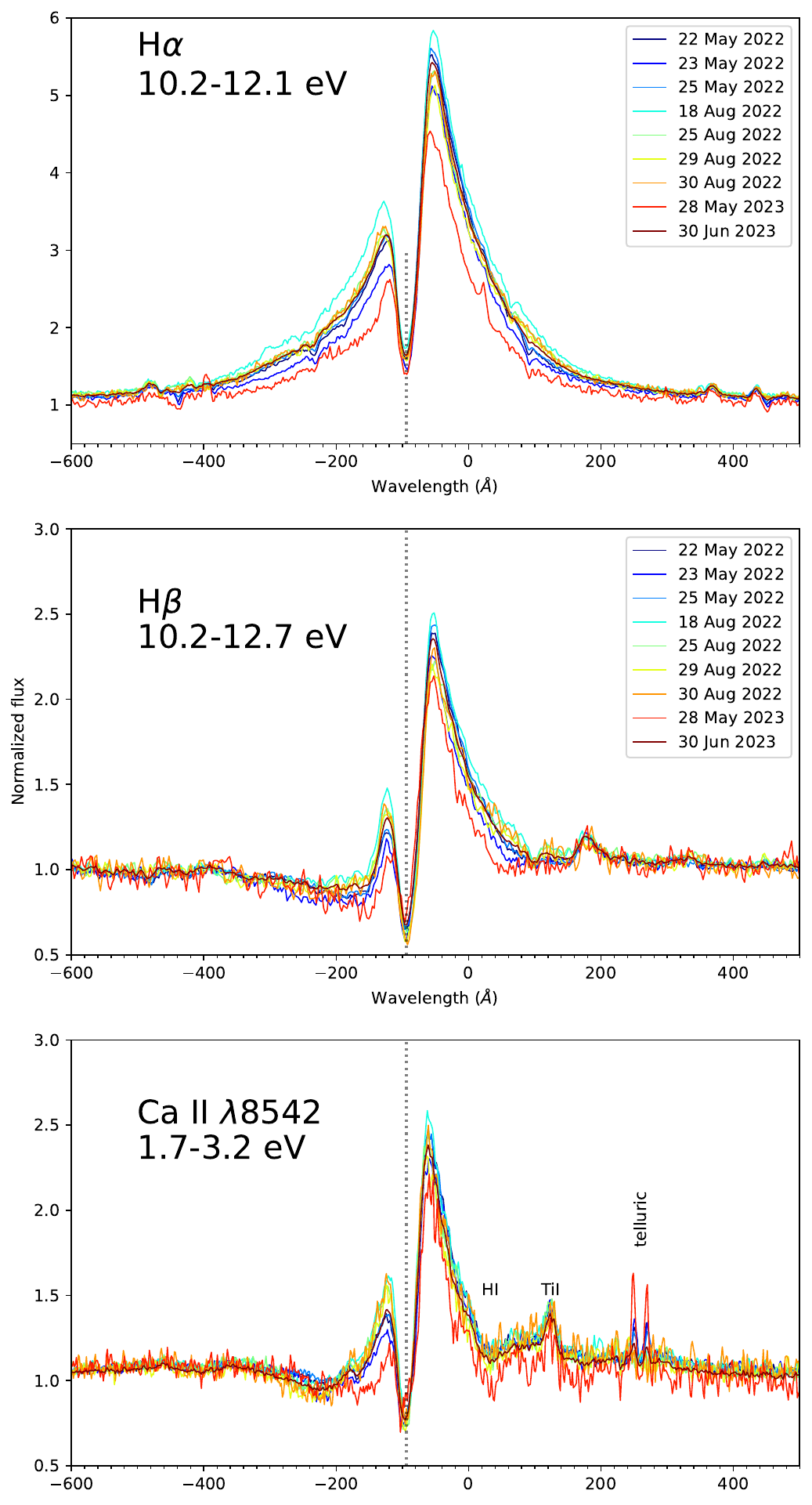}
    \caption{Profiles of selected lines. The dotted vertical lines mark the center of the narrow absorption component discussed in the text.}
    \label{fig-profiles-Ha}
\end{figure}

\begin{figure}
    \centering
    \includegraphics[width=\columnwidth]{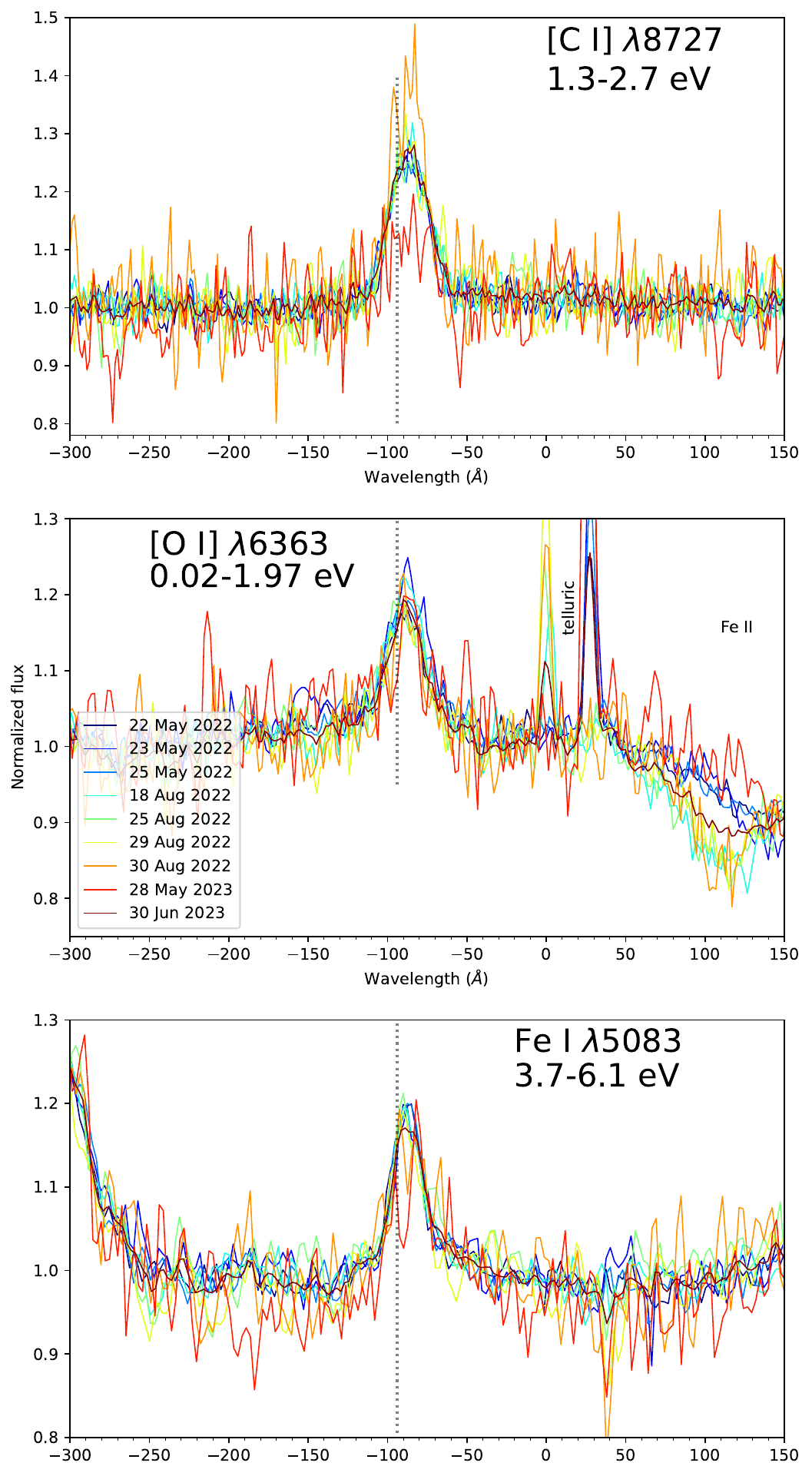}
    \caption{As Fig.\,\ref{fig-profiles-Ha}.}
    \label{fig-profiles-CI}
\end{figure}

\begin{figure}
    \centering
    \includegraphics[width=\columnwidth]{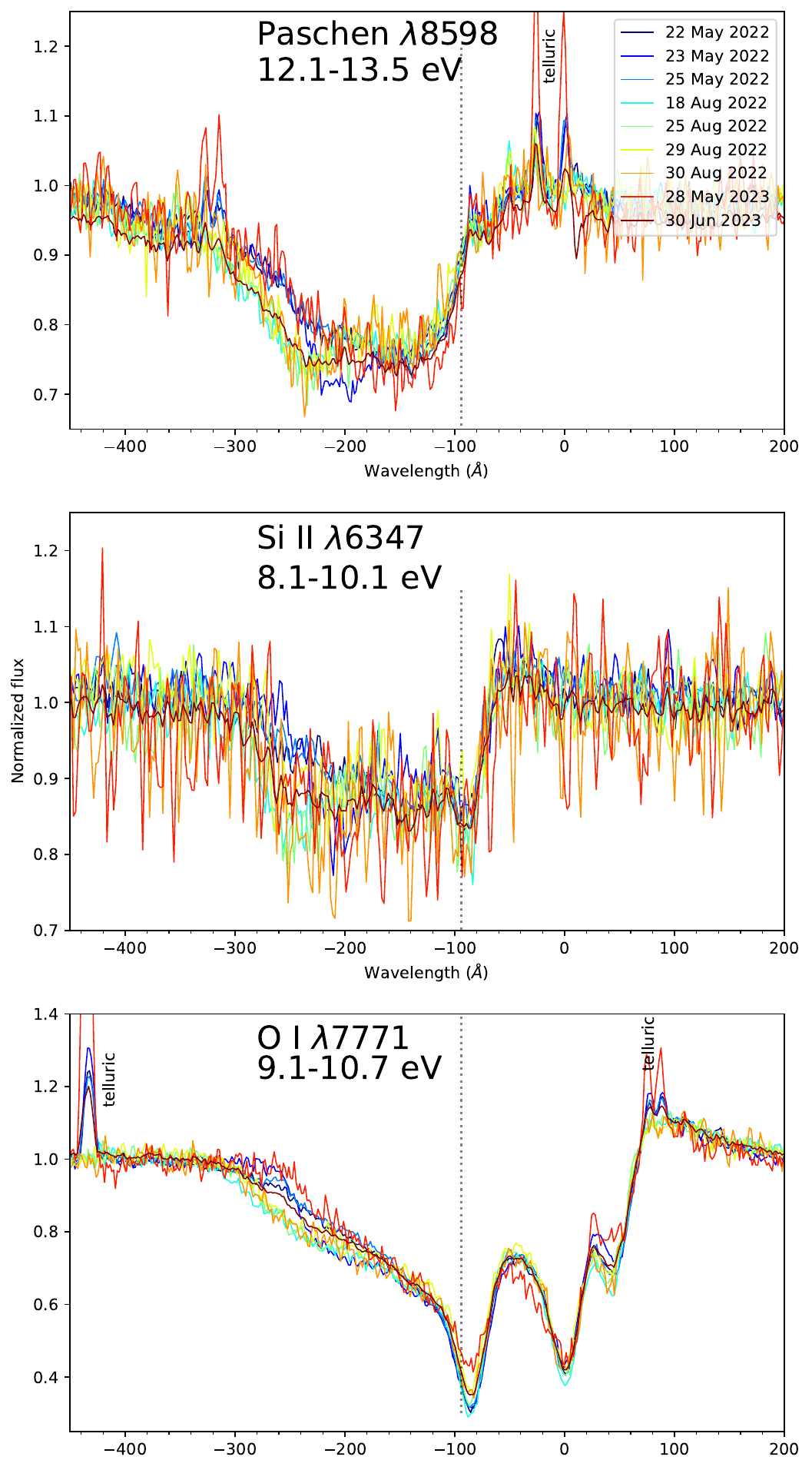}
    \caption{As Fig.\,\ref{fig-profiles-Ha}.}
    \label{fig-profiles-Paschen}
\end{figure}

\begin{figure}
    \centering
    \includegraphics[width=\columnwidth]{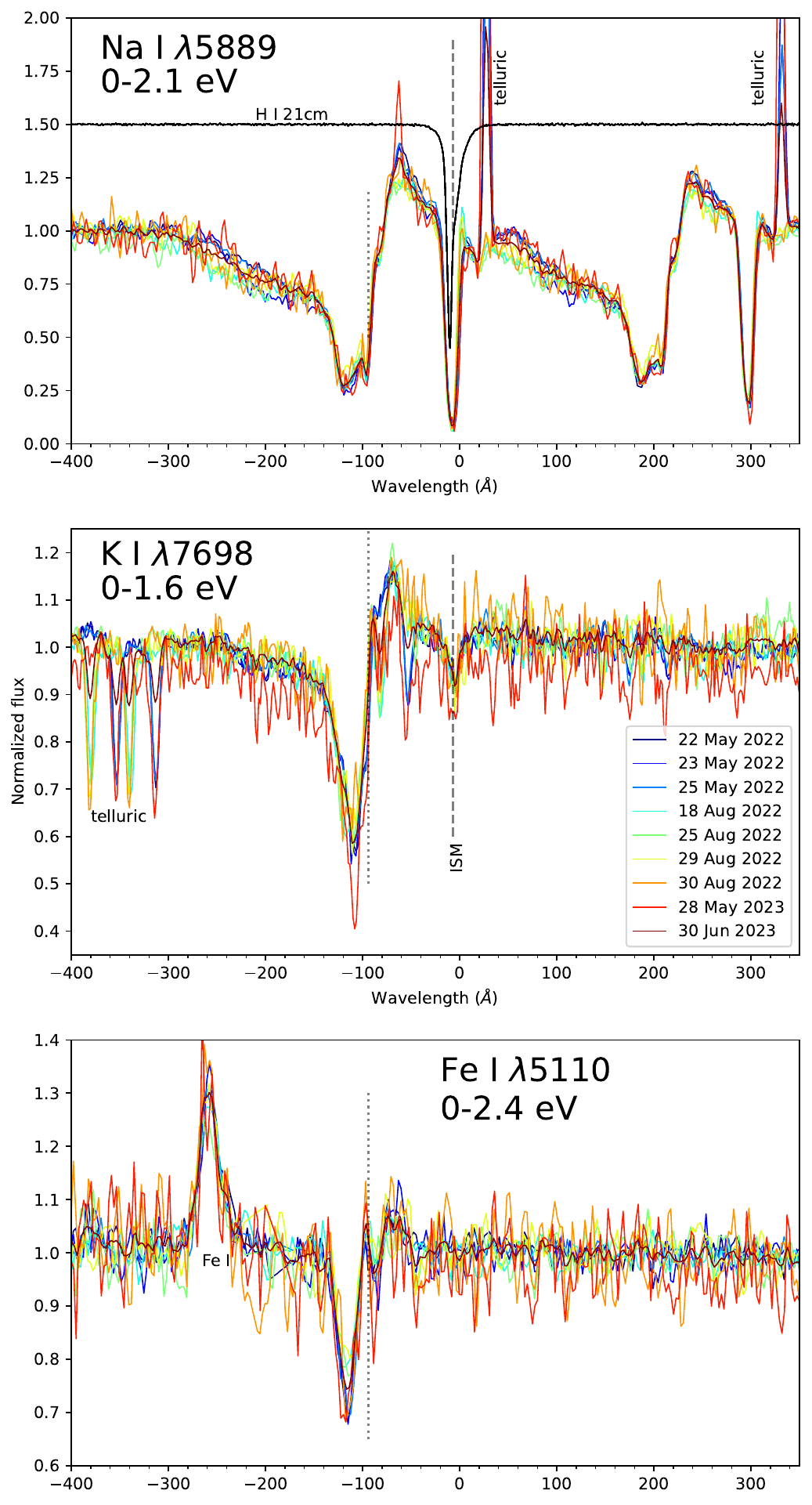}
    \caption{As Fig.\,\ref{fig-profiles-Ha}. Additionally, the scaled and inverted profile of the \ion{H}{I} emission line at 21 cm is shown with a black line. The interstellar component is marked with a dashed line.}
    \label{fig-profiles-Na}
\end{figure}

\begin{figure}
    \centering
    \includegraphics[width=\columnwidth]{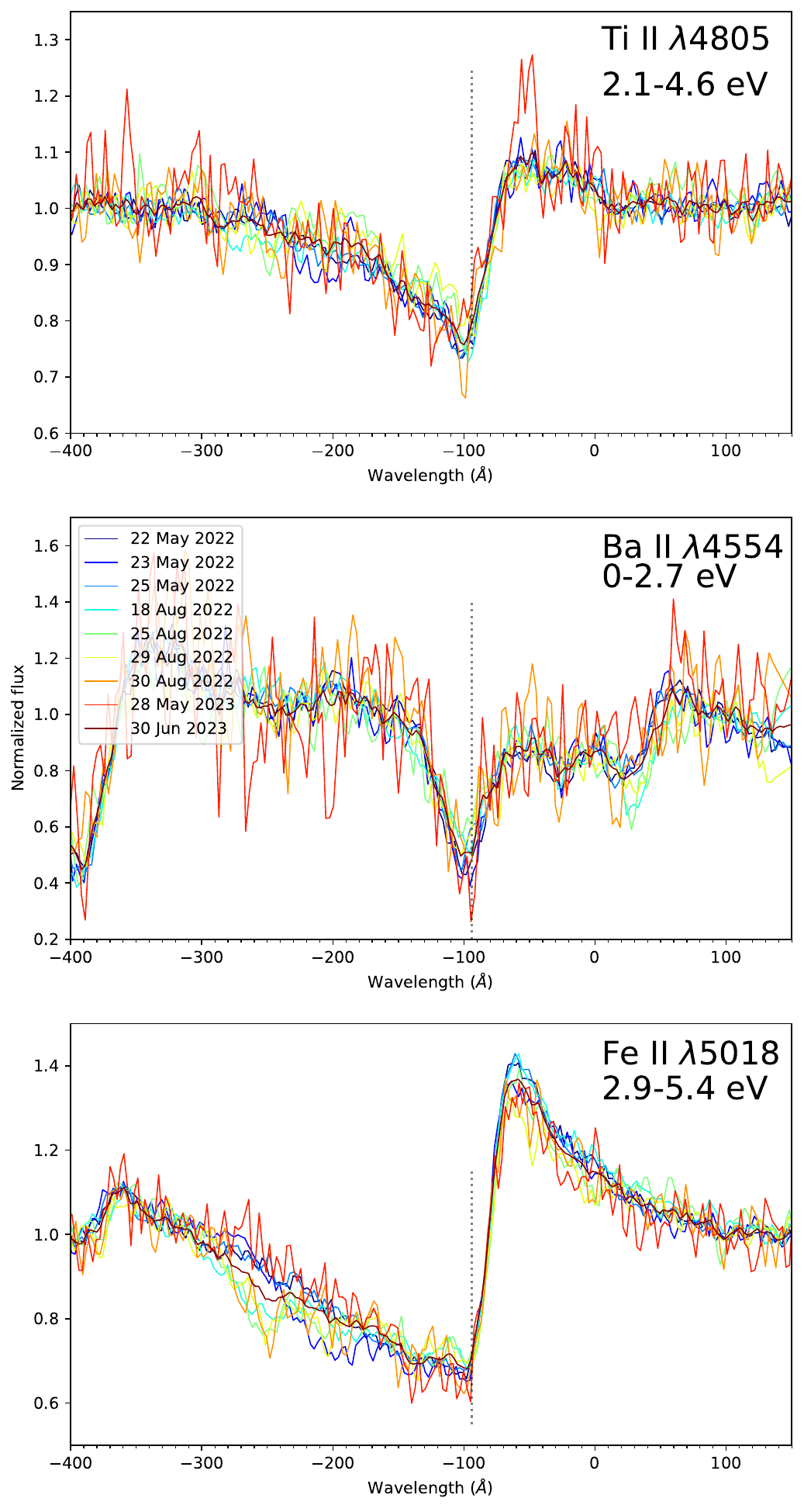}
    \caption{As Fig.\,\ref{fig-profiles-Ha}.}
    \label{fig-profiles-TiII}
\end{figure}

\end{appendix}

\end{document}